\documentclass[prb,aps,twocolumn,floatfix,amsmath,superscriptaddress]{revtex4}

\usepackage{graphicx}
\usepackage{color}

\begin{document}

\title{Anisotropic Pyrochlores and the Global Phase Diagram of the
   Checkerboard Antiferromagnet}

\author{Oleg A. Starykh}
\affiliation{Department of Physics, University of Utah,
Salt Lake City, UT 84112-0830}

\author{Akira Furusaki}
\affiliation{Condensed Matter Theory Laboratory, RIKEN,
Wako, Saitama 351-0198, Japan}

\author{Leon Balents}
\affiliation{Department of Physics, University of
California, Santa Barbara, CA 93106-4030}


\date{April 28, 2005}

\begin{abstract}

   We study the phase diagram of two models of spin-$1/2$
   antiferromagnets composed of corner-sharing tetrahedra, the basis of
   the pyrochlore structure.  Primarily, we focus on the Heisenberg
   antiferromaget on the checkerboard lattice (also called the planar
   pyrochlore and crossed-chains model).  This model has an anisotropic
   limit, when the dimensionless ratio of two exchange constants,
   $J_\times/J \ll 1$, in which it consists of one-dimensional spin
   chains coupled weakly together in a frustrated fashion.  Using
   recently developed techniques combining renormalization group ideas
   and one-dimensional bosonization and current algebra methods, we
   show that in this limit the model enters a {\sl crossed dimer} state
   with two-fold spontaneous symmetry breaking but no magnetic order.
   We complement this result by an approximate ``quadrumer triplet
   boson'' calculation, which qualitatively captures the physics of the
   ``plaquette valence bond solid'' state believed to obtain for
   $J_\times/J \approx 1$.  Using these known points in parameter
   space, the instabilities pointed to by the quadrumer boson
   calculation, and the simple limit $J_\times/J \gg 1$, we construct a
   few candidate global phase diagrams for the model, and discuss the
   nature of the quantum phase transitions contained therein.  Finally,
   we apply our quasi-one-dimensional techniques to an anisotropic
   limit of the {\sl three-dimensional} pyrochlore antiferromagnet, an
   approximate model for magnetism in GeCu$_2$O$_4$.  A crossed dimer
   state is predicted here as well.

\end{abstract}
\maketitle

\section{Introduction}
\label{intro}

\begin{figure}
\center
\includegraphics[width=0.7\columnwidth]{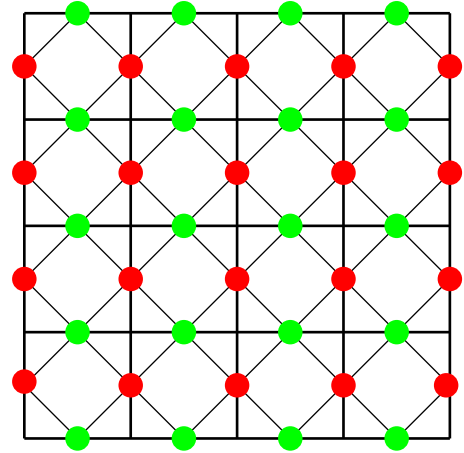}
\caption{(Color online)
  Heisenberg antiferromagnet on the checkerboard lattice, viewed as
  coupled spin chains.  Horizontal (vertical) spin chains run along
  the $x$ ($y$) axis.  Spins belonging to the horizontal (vertical)
  chains are shown as green (red) filled circles.  The intra-chain
  exchange (thick lines) is $J$, and the inter-chain exchange (diagonal
  thin lines) is $J_\times$.  }
\label{ccm-pic}
\end{figure}

We consider one of the most frustrated two-dimensional (2D)
antiferromagnets: the checkerboard antiferromagnet, also known as the
planar pyrochlore and the crossed-chains model (CCM).
As the name suggests, this model is motivated by the three-dimensional
(3D) pyrochlore materials. The 2D model is obtained by
a projection of 3D corner-sharing lattice of tetrahedra on a 2D plane.
This projection maps a four-spin tetrahedron onto a four-spin square
with additional links (antiferromagnetic exchanges) along the diagonals.
The structure obtained in this way, depicted in Fig.~\ref{ccm-pic},
preserves the corner-sharing arrangement of crossed squares, typical
of the original 3D pyrochlore lattice, but destroys the symmetry
between bonds of the tetrahedron: in two dimensions, the horizontal
and vertical bonds are not equivalent to diagonal ones. This lowering
of symmetry suggests consideration of extended 2D models with
the checkerboard structure where exchange interactions on
horizontal/vertical and diagonal bonds take on different values.
Among these, the quasi-one-dimensional limit, in which exchange along
horizontal and vertical directions $J$ is much stronger than that
along diagonal directions $J_\times$, is of special interest because
it involves competition between strong quantum fluctuations, typical
for one-dimensional (1D) spin chains, and equally strong geometric
frustration encoded in the structure of the crossed-chains lattice.

The resulting checkerboard antiferromagnet has been analyzed by a
variety of techniques along several complimentary ``directions" in the
parameter space: semi-classical analysis in the limit of large spin
$S\gg 1$,\cite{paradigm,canals,olegt} large-$N$ expansion,
\cite{moessner,sachdev,toronto} easy-axis generalization (of the 3D
model) \cite{hermele} and a quasi-1D ($J_\times/J \ll 1$)
approach.\cite{sliding}  In parallel with analytic approaches, the
model was investigated numerically via exact diagonalization studies
\cite{chalker,fouet,sindzingre} and cluster-based strong-coupling
expansion techniques.\cite{brenig,altman,brenig2}  The present paper
complements these approaches by combining a controlled analysis of the
quasi-1D limit with general arguments to pin down limits
of the phase diagram and postulate a likely global phase structure of
the model.

We begin by expounding the more general context of the problem.  One
of the central theoretical motivations behind the study of frustrated
quantum magnets is the hope that, when magnetic ordering is suppressed
by frustration, more novel types of order or even criticality may
emerge.  Phenomenological approaches suggest possible interesting
quantum phases exhibiting ``valence bond solid'' (VBS) order, in which
spins pair into singlets that are spontaneously localized on specific
bonds, breaking lattice symmetries.  More exotically, such approaches
suggest the possibility of phases with ``topological order'', in which spins
fluctuate quantum mechanically in a liquid-like state with however
subtle topological properties and often excitations with anomalous
(e.g., fractional) quantum numbers.  More recent predictions from such
theories also include ``deconfined'' quantum critical points and
phases in which several types of quasi-long-range (power-law) orders
coexist unconnected by microscopic symmetries.

Unfortunately, these types of phenomenological methods do not give
precise guidance as to the specific models in which such quantum
orders appear, and attempts to find them in realistic microscopic
Hamiltonians have met with at best limited success.  The one specific
context in which examples of all the above phenomena are, however, known
to occur is in one-dimensional spin chains.  Moreover, the theoretical
and microscopic understanding of such spin models is vastly more
complete than in two or three dimensions.  A natural hunting ground
for the exotic phenomenology described above would hence seem to lie
in spin models consisting of chains weakly coupled into two or
three dimensional arrays.  A recently gained understanding of the
crucial role of nominally irrelevant operators and
fluctuation-generated interactions in describing frustrated quasi-1D
magnetic systems,\cite{oleg-leon} described below, brings the hunt to
(some degree of) fruition.

In this paper, as in a previous work,\cite{oleg-leon} we
follow this approach, taking as the weakly coupled units
in question $S=1/2$ Heisenberg nearest-neighbor antiferromagnetic
chains (other further-neighbor interactions along each chain may
be included, provided they are not overly strong).  A cause for hope is
that such a 1D chain is well known to exhibit a critical ground state
with power-law correlations of various types.  One prominent type of
correlation in such a chain is antiferromagnetic, specifically:
\begin{equation}
      \label{eq:antiferro}
      \langle \vec{S}(n)\cdot \vec{S}(n')\rangle \sim
      \frac{(-1)^{n-n'}}{|n-n'|} + \cdots,
\end{equation}
where $n$ is the coordinate along the chain, and the brackets indicate
a ground state expectation value.  The omitted $\cdots$ terms decay
much faster ($\sim 1/|n-n'|^2$ or faster) than the dominant
slowly-decaying antiferromagnetic one shown here (we have also for
simplicity neglected an unimportant multiplicative logarithmic
correction to this term).  The dominance of antiferromagnetic
correlations in the two-spin correlation function often leads to the
misconception that a good picture of the ground state of the 1d
Heisenberg chain is that of fluctuating local antiferromagnetic order,
i.e., a magnet in which spins are locally N\'eel ordered but the
quantization axis fluctuates in space and time.  Such a picture is in
fact incomplete.  This becomes clear upon considering the fluctuation
of the local bond energy or dimerization,
\begin{equation}
      \label{eq:bonden}
      B(n) = \vec{S}(n)\cdot\vec{S}(n+1)
             - \langle\vec{S}(n)\cdot\vec{S}(n+1)\rangle.
\end{equation}
One finds that its (staggered) correlations,
\begin{equation}
      \label{eq:bondcorr}
      \langle B(n) B(n')\rangle \sim  \frac{(-1)^{n-n'}}{|n-n'|} + \cdots,
\end{equation}
have {\sl precisely} the same slow power-law decay 
(again, up to a 
multiplicative logarithmic correction) as the
antiferromagnetic ones in Eq.~(\ref{eq:antiferro}) above!  Further
examination of other correlators reveals no additional power-law
correlations with competitive slow decay. Thus the 1D
Heisenberg chain should be thought of as consisting of locally
fluctuating antiferromagnetic {\sl and} valence bond solid order of
comparable strength.

With this understanding, it is natural to expect that weakly-coupled
arrays of such chains might be pushed by the inter-chain coupling into
magnetically ordered, dimer ordered, or perhaps critical states, if
this coupling favors the intrinsic antiferromagnetic or VBS ordering
tendency, or fosters their balanced competition, respectively.  While
we believe this reasoning to be essentially correct, for many years,
the richness of such possible behaviors went unrealized in the
literature.  This is because if the spin chains are linked by magnetic
two-spin Heisenberg interactions, these couple primarily to the
antiferromagnetic fluctuations within the chains, and not to the VBS
ones.  Hence, for such a case, the problem of Heisenberg spin chains
coupled by weak inter-chain interactions is rather well understood.
With non-frustrated transverse (with respect to the chain direction)
couplings, both renormalization group \cite{rg-affleck} and
self-consistent mean-field analysis \cite{schulz} predict an instability
towards classical long-range ordered phase characterized by a non-zero
expectation value of the spin $\langle \vec{S}_r\rangle \neq 0$.  This
instability follows from the correlations in
Eq.~(\ref{eq:antiferro}), which, loosely speaking, make the spin chain
highly susceptible to magnetic ordering.

More recently, it was recognized that the situation becomes more
interesting and less clear-cut when the inter-chain interaction is
strongly frustrated, as is the case for the crossed-chains
model we investigate here. The effect of frustration is to reduce and,
ultimately, nullify the effective inter-chain magnetic field
experienced by spins of the chain due to transverse inter-chain
exchange interactions (due to cancellations between contributions from
spins whose local fluctuating orientations, according to
Eq.~(\ref{eq:antiferro}), are antiparallel). With no effective external
field present, the classical ordering instability is naturally absent,
resulting in (almost) decoupled behavior of distinct spin chains.  In formal
calculations embodying this physical picture, the weak residual
inter-chain interaction which does not cancel with predominantly
antiferromagnetic correlations appears to be described by the scalar product
of conserved spin currents from the chains involved. This observation
led to the proposal that, as a result, the system of such
coupled chains forms a liquid-like ground state with fractionalized
spin excitations (spinons). The systems considered included a frustrated
spin ladder,\cite{allen-essler-ners,kim} its 2D extension, i.e.,
the spatially-anisotropic frustrated square lattice
antiferromagnet,\cite{NT} and the crossed-chains model.\cite{sliding}

As shown in Ref.~\onlinecite{oleg-leon}, 
in the former two cases these
conclusions are in fact incorrect, due to the neglect of the VBS
correlations, Eq.~(\ref{eq:bondcorr}), equally as inherent as the
antiferromagnetic ones to the Heisenberg chain.  Although the
microscopic magnetic exchange between spins on different chains does
not directly couple to VBS fluctuations, such a dimer coupling between
(certain pairs of) chains is inevitably generated by the weak
residual magnetic interactions remaining after the dominant
antiferromagnetic cancellation.  A careful analysis of the types of
such dimer couplings allowed by symmetry and the detailed mechanism of
their generation are crucial in determining the fate of the spin
system and the strength of any ordering tendency.

Technically, this analysis can be accomplished in a controlled fashion
using powerful field-theoretical methods borrowed from 1D physics.
The point, made in Ref.~\onlinecite{oleg-leon}, is that no
fine-tuning of the two-spin inter-chain
exchange interaction can make the low-energy field theory
{\em exactly} of current-current type.  Some higher-order derivative
terms (typically involving spatial derivatives of the staggered
magnetization field) are bound to be present (as getting rid of all of
them to all orders would require tuning infinite number of inter-chain
couplings to zero). Such derivative terms are commonly neglected on
the grounds of their irrelevance with respect to the Luttinger
liquid fixed point of the independent spin chain. However, the
quasi-1D problem is not the same as the purely
1D one.    Instead of disregarding irrelevant
high-derivatives terms from the outset, one has to consider if they,
in combination with the leading current-current term, can produce
quantum corrections to the {\em relevant} inter-chain couplings.  This
indeed occurs both in the models of Ref.~\onlinecite{oleg-leon}, and,
as we will see, in the crossed-chains model studied here.

\begin{figure}
\center
\includegraphics[width=0.7\columnwidth]{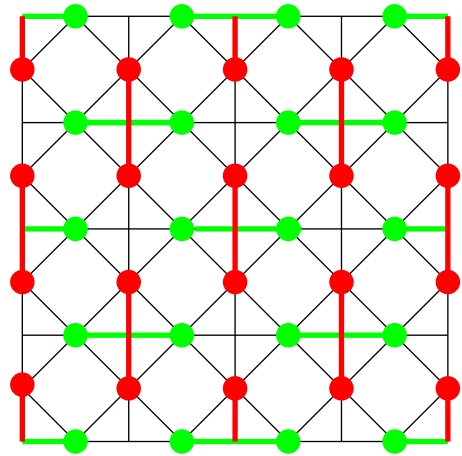}
\caption{(Color online) 
Crossed-dimer dimerization pattern. 
``Strong'' bonds (ones
where $\epsilon > 0$)
on horizontal (vertical) chains are shown in green (red). As before,
spins on horizontal (vertical) chains
are denoted by green (red) circles.}
\label{patterns}
\end{figure}

In the present paper we extend the analysis of Ref.~\onlinecite{oleg-leon}
to the CCM and show that previous claim of the sliding
Luttinger liquid ground state \cite{sliding} is not correct. Instead,
similarly to the spatially-anisotropic square lattice model discussed
above, the ground state is of spontaneously dimerized type, albeit
with staggered ordering of dimers on parallel chains. The resulting
configuration, shown in Fig.~\ref{patterns}, can be described as a
{\sl crossed-dimer} one.

The paper is organized as follows.  Section~\ref{latticeH} describes
the Hamiltonian of the CCM model, its lattice symmetries, and the
passage to the field-theoretical description of the low-energy degrees
of freedom and the operator product algebra they form.  Section
\ref{sec:solution} describes perturbative analysis of the model in the
one-dimensional limit of weakly coupled chains, $J_\times/J \ll 1$. It
contains key technical details of our work and explains the mechanism
by which the crossed-dimer phase is stabilized.  The limit of the
fully two-dimensional model ($J_\times\approx J$), the planar
pyrochlore antiferromagnet, is analyzed within the plaquette-operator
mean-field approximation in Sec.~\ref{planar-pyro}.  This is followed
by Sec.~\ref{global}, which summarizes the preceding material in terms of
two possible scenarios for the global zero-temperature phase
diagram of the checkerboard antiferromagnet. There we present
phenomenological symmetry-based analyses of the quantum phase
transitions between various phases of the model (and also point out an
interesting connection with the recent deconfined quantum critical
point idea). Section~\ref{sec:3d} describes a three-dimensional
extension of our model, the quasi-one-dimensional pyrochlore
antiferromagnet, and its possible relevance to the experiments on
GeCu$_2$O$_4$ and ZnV$_2$O$_4$. Our main points are briefly summarized
in Sec.~\ref{sec:conclusions}.  Two Appendices contain important
technical details of the fermionic formulation of the low-energy
sector of the $S=1/2$ isotropic Heisenberg chain.

\section{From Lattice to Continuum Field Theory}
\label{latticeH}

\subsection{Lattice model and symmetries}
\label{sec:latt-model-symm}

The Hamiltonian of the system $H$ describes a collection of horizontal
($H_h$) and vertical ($H_v$) Heisenberg chains interacting with each
other via the inter-chain interaction $V$:
\begin{equation}
    H=H_0 + V= H_h + H_v + V.
\label{lattice-H}
\end{equation}
Spins ($S=1/2$) are located at the sites of the checkerboard
(crossed-chains) lattice shown in Fig.~\ref{ccm-pic}.  The crossings
of the lattice have integer coordinates $(n,m)$, so the sites
of horizontal chains have half-integer $x$-coordinates $n+\frac12$ and
integer $y$-coordinate $m$, while sites of the vertical chains are
described by $(n,m+\frac12)$ pairs. With this convention the Hamiltonian
of horizontal chains reads
\begin{equation}
    H_h= J \sum_{n,m} \vec{S}_h(n-1/2,m)
\cdot \vec{S}_h(n+1/2,m).
\label{lattice-H-horiz}
\end{equation}
Similarly, $H_v$ is given by
\begin{equation}
H_v= J \sum_{n,m} \vec{S}_v(n,m-1/2) \cdot \vec{S}_v(n,m+1/2).
\label{lattice-H-vert}
\end{equation}
With local uniform magnetization defined by
\begin{subequations}
\begin{eqnarray}
\vec{s}_h(n,m)&=&
\vec{S}_h(n-1/2,m)+\vec{S}_h(n+1/2,m),
\qquad\\
\vec{s}_v(n,m)&=&
\vec{S}_v(n,m-1/2)+\vec{S}_v(n,m+1/2),
\end{eqnarray}
\label{J_h J_v}
\end{subequations}
the inter-chain interaction reads
\begin{equation}
V=J_\times\sum_{n,m}\vec{s}_h(n,m)\cdot\vec{s}_v(n,m)
\label{lattice-V}
\end{equation}
and is characterized by the inter-chain exchange $J_\times > 0$
which is much smaller than the in-chain antiferromagnetic exchange
$J > 0$.
We note that $J_\times$ is the nearest-neighbor exchange on the
checkerboard lattice while $J$ is the next-nearest-neighbor
exchange interaction.

The space group symmetries of $H$, Eq.~(\ref{lattice-H}), can now be
summarized.  The translational subgroup is generated by unit
translation along the horizontal chains $T_h$ and that along the
vertical chains $T_v$.  The remainder
is generated by $\pi/2$ rotations about a crossing, and reflections
about e.g., a vertical line through either a site or midpoint of a bond
of a horizontal chain.  We denote these two operations ``site parity''
$P_{sh}$ and ``link parity'' $P_{Lh}$, respectively.  As these are
microscopic lattice symmetries, they will be preserved by any
renormalization group transformation.  Observe that $P_L$ is a product
of two other operations: $P_{Lh}=P_{sh} \circ T_h$.

\subsection{Continuum field theory and scaling operators}
\label{subsec:CFT}

The limit $J_\times \ll J$ allows us to approach the problem
from one-dimensional perspective: we treat $V$ as a perturbation and
ask whether it can destabilize the critical ground state of the
independent (decoupled) spin chains.  The smallness of the
$J_\times/J$ ratio allows us to take the continuum limit along every chain
involved.  As mentioned in the Introduction, a single Heisenberg chain
is described in the continuum limit (i.e., at low energies) by a
universal critical theory, with a variety of power-law correlations.
Formally, this is most compactly described as the Wess-Zumino-Witten
(WZW) $SU(2)_1$ theory \cite{wzw}, with the action (in $1+1$ 
dimensions) \cite{itzykson,gnt-book}
\begin{eqnarray}
      \label{eq:WZW}
      S_{WZW} & = & \frac{1}{8\pi} \int\! d^2x \, {\rm Tr}\, \partial_\mu
      g^\dagger \partial_\mu g \nonumber \\
&&  - \frac{i}{12\pi}\int\! d^3 x\, \epsilon_{\mu\nu\lambda} {\rm Tr}\,
      g^\dagger \partial_\mu g g^\dagger \partial_\nu g g^\dagger
      \partial_\lambda g .
\end{eqnarray}
Here $g$ is an $SU(2)$ matrix. The coordinate $x_0=v\tau$ ($v$ is the
spin velocity and $\tau$ is imaginary time) and $x_1=x$,
the coordinate along the chain, and $d^3x$ is defined by extending
this 2D space into a three-dimensional hemisphere $x_2<0$, 
the boundary of which is the (compactified) physical 2D plane $(x_0,x_1)$,
and analytically continuing $g(x_0,x_1)$ into this hemisphere such that
$g(x_0,x_1,x_2 \to -\infty)\rightarrow 1$ and $g(x_0,x_1,x_2=0)=g(x_0,x_1)$.  This formal action is
not of very much direct practical use, but serves to illustrate the
underlying degrees of freedom of the critical theory.  All operators
in the WZW $SU(2)_1$ theory can be constructed from $g$.  Corresponding
to the two dominant power-law correlations in
Eqs.~(\ref{eq:antiferro}) and (\ref{eq:bondcorr}), there are two scaling
operators \cite{itzykson,gnt-book}
\begin{eqnarray}
      \label{eq:domops}
      \vec{N} & \sim & -i \, {\rm Tr}\, g\vec\sigma, \\
      \epsilon & \sim & {\rm Tr}\, g.
\end{eqnarray}
Here $\vec\sigma$ is the vector of Pauli matrices, and the $\sim$
indicates that the proportionality between these fields and the
physical staggered magnetization/dimerization involves a cut-off
dependent factor.  The operator $\vec{N}$ represents the local
staggered magnetization, while $\epsilon$ represents the local
staggered dimerization (it is the continuum version of the bond 
operator 
in (\ref{eq:bonden})).  There are also subdominant power-law
correlations arising from fluctuations of the chiral $SU(2)$ currents,
\begin{eqnarray}
      \label{eq:currents}
      \vec{J}_R & = & \frac{1}{4\pi}{\rm Tr}\, g^\dagger \bar\partial  g
      \vec\sigma, \\
      \vec{J}_L & = & \frac{1}{4\pi}{\rm Tr}\,  g \partial  g^\dagger
      \vec\sigma,
\end{eqnarray}
with $\partial = (\partial_0 - i\partial_1)/2$, and $\bar\partial =
(\partial_0 + i\partial_1)/2$.  Physically, the operator $\vec{J}=\vec{J}_R
+ \vec{J}_L$ represents the local uniform magnetization, while
$v(\vec{J}_R-\vec{J}_L)$ represents the local magnetization (spin
transport) current.

All the low-energy power-law correlations of the weakly coupled
Heisenberg chains can be exposed by decomposing lattice operators into
a set of the above continuum operators (and generally their
derivatives, see below) {\sl for each chain}.  This, for example,
leads to the following decomposition of the spin at a site $n-1/2$
along the horizontal chain number $m$:
\begin{equation}
\vec{S}_{h}(n-1/2,m)=
a\left[\vec{J}_{h,m}(x) + (-1)^n \vec{N}_{h,m}(x)\right].
\label{spin-operator}
\end{equation}
Here $x=na$ ($a$ is the lattice spacing) and $\vec{J}$
(respectively, $\vec{N}$) represents uniform (respectively, staggered)
part of the spin density.
Similarly, for the vertical spin chains we have
\begin{equation}
\vec{S}_v(n,m-1/2)=
a\left[\vec{J}_{v,n}(y) + (-1)^m \vec{N}_{v,n}(y)\right],
\end{equation}
where $y=ma$.
Notice that the
continuum limit is taken only for the  coordinate along the chain; the
perpendicular one becomes an index $m$ (respectively, $n$ for vertical
chains).
The uniform spin magnetization $\vec{J}$ is the sum of the
right $(\vec{J}_R)$ and left $(\vec{J}_L)$ moving components,
$\vec{J}=\vec{J}_R + \vec{J}_L$, and represents the conserved spin
density (it is often referred to in the literature as the spin
``current'', the term originating from the relativistic concept of
space-time current, whose time component is the conserved density).
Note that the staggered dimerization $\epsilon$ does not appear in
Eq.~(\ref{spin-operator}); in fact, it cannot appear in the decomposition
of any single spin operator since it is not a vector under $SU(2)$.
As discussed in the Introduction, for
this reason dimer order does not appear likely in weakly coupled
Heisenberg chains with unfrustrated inter-chain couplings.

The action of the microscopic space group symmetries (described above)
upon the continuum scaling operators will be crucial in the
following.  These are rather clear on physical grounds \cite{eggert}:

{\em Translation:}
\begin{equation}
T:
\vec{J} \to \vec{J}, \quad
\vec{N} \to -\vec{N}, \quad
\epsilon \to -\epsilon.
\label{Tr}
\end{equation}

{\em Site parity:}
\begin{equation}
P_s:
\vec{J} \to \vec{J}, \quad
\vec{N} \to \vec{N}, \quad
\epsilon \to -\epsilon.
\label{P_s}
\end{equation}

{\em Link parity:}
Using $P_L=P_s \circ T$ we find
\begin{equation}
P_L:
\vec{J} \to \vec{J}, \quad
\vec{N} \to -\vec{N}, \quad
\epsilon \to \epsilon.
\label{P_L}
\end{equation}
We will see at the end of this section that this symmetry is
responsible for the absence of $\vec{N}_v\cdot \vec{N}_h$ terms in the
Hamiltonian of the problem.

Because of somewhat non-intuitive point-splitting identities, the WZW
model can be written in Hamiltonian form (known as the Sugawara form)
in terms of the spin currents.  For a single chain, one has
\begin{equation}
      H_{WZW} = \frac{2\pi v}{3} \int\! dx\,
    \left[ \vec{J}_R(x)\cdot\vec{J}_R(x) +
      \vec{J}_L(x)\cdot\vec{J}_L(x)\right] .
\label{eq:sugawara}
\end{equation}
Applied to the set of horizontal chains (labelled by $m$), the lattice
Hamiltonian $H_h$, Eq.~(\ref{lattice-H-horiz}), transforms into
\begin{eqnarray}
H_h&=&\frac{2\pi v}{3}\sum_m \int dx
\left[ \vec{J}_{h,m,R}(x)\cdot\vec{J}_{h,m,R}(x)
\right.
\nonumber\\
&&
\qquad\qquad\qquad{}
+\vec{J}_{h,m,L}(x)\cdot\vec{J}_{h,m,L}(x)
\nonumber\\
&&\qquad\qquad\qquad
\left.{}
    + g_{bs} \vec{J}_{h,m,R}(x)\cdot\vec{J}_{h,m,L}(x)
\right].
\qquad
\label{H-horiz}
\end{eqnarray}
Here $v = \frac{\pi}{2} J a$ is the spin velocity.  Note again that
$J^a_{h,m,R/L}(x)$ depends on position $x=n a$ along the chain
direction whereas its $y=m a$ coordinate dependence only shows up via
the (horizontal) chain index $m$.  We have actually included in
Eq.~(\ref{H-horiz}) a  {\sl correction} (proportional to $g_{bs}$)
to the WZW model, which is present
in the Heisenberg chain but is {\sl marginally irrelevant} in the
situation under consideration.  For this reason, it may be safely
neglected in what follows.
Similarly
\begin{eqnarray}
H_v&=&\frac{2\pi v}{3}\sum_n \int dy
\left[ \vec{J}_{v,n,R}(y)\cdot\vec{J}_{v,n,R}(y)
\right.
\nonumber\\
&&\qquad\qquad\qquad{}
+\vec{J}_{v,n,L}(y)\cdot\vec{J}_{v,n,L}(y)
\nonumber\\
&&\qquad\qquad\qquad
\left.{}
+ g_{bs} \vec{J}_{v,n,R}(y)\cdot\vec{J}_{v,n,L}(y)
\right].
\qquad
\label{H-vert}
\end{eqnarray}

\subsection{Decomposition of the full lattice model}

Now we are ready to express the inter-chain perturbation Eq.~(\ref{lattice-V})
in terms of low-energy modes $\vec{J}$ and $\vec{N}$.  We begin by
analyzing the sum of two neighboring spins on the same (say,
horizontal) chain,
\begin{eqnarray}
\vec{s}_h(n,m)&=&
\vec{S}_h(n-1/2,m)+\vec{S}_h(n+1/2,m)
\nonumber\\
&=&
a\left[2\vec{J}_{h,m}(x) - (-1)^n a\partial_x \vec{N}_{h,m}(x)\right].
\quad
\end{eqnarray}
For the reasons to be explained in detail below, we have retained the
next-to-leading irrelevant contribution $(\partial_x \vec{N})$ in this
expression. Similar decomposition is done for the sum of two spins on
the crossing vertical chain. The interchain interaction $V$ thus
reads
\begin{eqnarray}
V&=&\sum_{n,m}\left\{g_{jj}\vec{J}_{h,m}(x)\cdot\vec{J}_{v,n}(y)
\right.
\nonumber\\
&&\qquad{}
    -g_{nj}\left[(-1)^n \partial_x\vec{N}_{h,m}(x)\cdot\vec{J}_{v,n}(y)
\right.
\nonumber\\
&&\left.\qquad\qquad{}
+ (-1)^m\vec{J}_{h,m}(x)\cdot\partial_y \vec{N}_{v,n}(y)\right]
\nonumber\\
&&\left.\qquad{}
+ g_{nn}(-1)^{n+m}\partial_x\vec{N}_{h,m}(x)\cdot\partial_y\vec{N}_{v,n}(y)
\right\},
\qquad
\label{V-ccm-full}
\end{eqnarray}
where, as before, $x=na, y=ma$ and the following couplings are
introduced to shorten notations:
\begin{equation}
g_{jj}=4J_\times a^2, \quad
g_{nj}=2J_\times a^3, \quad
g_{nn}=J_\times a^4.
\label{couplings}
\end{equation}
It is important to observe that Eq.~(\ref{V-ccm-full}) does not contain
$\vec{N}_h \cdot \vec{N}_v$ type of terms, which are forbidden by the
symmetry of the checkerboard lattice. For example, reflection with
respect to the vertical chain changes sign of $\vec{N}_h$,
$(P_L: \vec{N}_h \rightarrow -\vec{N}_h)$, while leaving $\vec{N}_v$
invariant, see Eq.~(\ref{P_L}).
This reflects strong frustration of the
model under study, as discussed in
the Introduction.  Observe also that any pair of horizontal and
vertical chains cross only once, which makes Eq.~(\ref{V-ccm-full}) {\em
      local} in space.  This requires us to think carefully about the
short-distance regularization of the low-energy theory defined by
Eq.~(\ref{H-horiz}), Eq.~(\ref{H-vert}), and Eq.~(\ref{V-ccm-full})
---the corresponding analysis is described in the next Section.

\subsection{Operator product expansion}

Various perturbations to the WZW model Eq.~(\ref{eq:sugawara}) [such
as the intra-chain backscattering $g_{bs}$ in Eq.~(\ref{H-horiz}) and
Eq.~(\ref{H-vert}), and the inter-chain $V$, Eq.~(\ref{V-ccm-full})]
are most conveniently analyzed with the help of {\sl operator product
   expansions} (OPE).  These are operator identities that are derived
by applying Wick's theorem to a correlation function of a pair of
operators at nearby points, say, $(x,\tau)$ and $(0,0)$ ---several of
the examples below are worked out in Appendix~\ref{a2}; see also
Appendix A of Ref.~\onlinecite{lin} for more examples.  The OPE below
are valid for operators from the same chain, and, to lighten
expressions, we suppress chain indices here.

The spin currents $\vec{J}_{R/L}$ obey the following chiral OPEs,
which are frequently used in the literature\cite{gnt-book} [these,
for example, are used to derive the renormalization-group flow of
$g_{bs}$ term in Eqs.~(\ref{H-horiz}) and (\ref{H-vert})]:
\begin{eqnarray}
J^a_R(x,\tau) J^b_R(0)&=&\frac{\delta^{ab}}{8\pi^2 v^2 (\tau - ix/v +
\alpha\sigma_\tau)^2}  \nonumber\\
&&+ \frac{i\epsilon^{abc}
J^c_R(0)}
{2\pi v (\tau - ix/v + \alpha\sigma_\tau)}
\label{ope-r}
\end{eqnarray}
and
\begin{eqnarray}
J^a_L(x,\tau) J^b_L(0)&=&\frac{\delta^{ab}}{8\pi^2 v^2 (\tau + ix/v +
\alpha\sigma_\tau)^2}  \nonumber\\
&&+ \frac{i\epsilon^{abc}
J^c_L(0)}
{2\pi v (\tau + ix/v + \alpha\sigma_\tau)},
\label{ope-l}
\end{eqnarray}
where, as explained in Appendix~\ref{app:Green-fermions},
$\alpha=a/v$ is the short-time cutoff of the theory
and $\sigma_\tau=\mathrm{sign}(\tau)$.

Being a conserved current, $J^a$ is also a generator of rotations.
Thus the OPE of $J^a$ and $N^a$ should be
nontrivial. In fact, this one is the most important OPE for the
subsequent analysis (see  Appendix~\ref{a2} for the derivation),
\begin{eqnarray}
J^a_R(x,\tau)N^b(0)&=&\frac{i\epsilon^{abc}N^c(0)
-i\delta^{ab}\epsilon(0)}{4\pi v(\tau-ix/v + \alpha\sigma_\tau)}
,\nonumber\\
J^a_L(x,\tau)N^b(0)&=&\frac{i\epsilon^{abc}N^c(0)
+i\delta^{ab}\epsilon(0)}{4\pi v(\tau+ix/v + \alpha\sigma_\tau)}.
\label{ope-JN}
\end{eqnarray}

Finally, fusing spin current with dimerization $\epsilon$ gives back
the staggered magnetization
\begin{eqnarray}
&&J^a_R(x,\tau) \epsilon(0)=
\frac{i N^a(0)}{4\pi v (\tau-ix/v + \alpha\sigma_\tau)} ,
    \nonumber\\
&&J^a_L(x,\tau) \epsilon(0)=
\frac{-i N^a(0)}{4\pi v (\tau+ix/v + \alpha\sigma_\tau)}.
\label{ope-Je}
\end{eqnarray}
Observe that Eqs.~(\ref{ope-r})--(\ref{ope-Je}) form a closed
operator algebra---this is the key technical reason behind the
generation of the inter-chain interaction of staggered
magnetizations
in frustrated spin chains models, see
Ref.~\onlinecite{oleg-leon} and Sec.~\ref{subsec:irrel} below.

\section{Low-energy Hamiltonian}
\label{sec:solution}
The spatially-anisotropic $J_1$-$J_2$ model \cite{oleg-leon} has
taught us that keeping track of the nominally irrelevant terms is
crucial for a correct solution of the problem. In this section, we
extend this line of thinking to the crossed-chains model and
demonstrate that indeed irrelevant terms produce symmetry-allowed
relevant ones in a simple perturbation theory.

\subsection{Symmetry analysis}
\label{sec:symmetry}
Before proceeding with microscopic calculations, it is instructive to
write down most general form of the inter-chain Hamiltonian $\delta V$
which is allowed by symmetries of the crossed-chains lattice.  The
reason to do so is that, while many such terms will be absent in a
na\"ive continuum limit of the original spin model, those which are
``accidentally'' missing (i.e. not prohibited by any symmetry) may be
expected to be generated as a ``quantum correction'' (i.e. through a
RG transformation) when na\"ively irrelevant terms are taken into
account.  The necessary complete set of space group generators for
this analysis, $T_h, P_{sh}, P_{Lh},$ and $R_{\pi/2}$, was introduced
in Sec.~\ref{sec:latt-model-symm}.

Naturally (as in any field theory), there are an infinite number of
possible interactions, and since there are additionally an infinite
number of chains, the multitude of potential terms is compounded.
Physically, however, ``pairwise'' interactions involving fields on
only two chains at a time are expected to be most important
(interactions involving more chains simultaneously can be shown to
occur only in higher order in $J_\times/J$).  Such an inter-chain
Hamiltonian naturally splits into the sum of $\delta V_\times$, which
describes interactions between two crossing chains, and $\delta
V_\parallel$, which includes interactions between {\em parallel}
chains, $\delta V = \delta V_\times + \delta V_\parallel$.  Within
these chain-pair interactions, we narrow the search by considering the
``most relevant'' possibilities (ones involving the smallest number of
the smallest scaling dimension primary fields $\vec{N}$ and $\epsilon$
and no derivatives).  Since we are perturbing the decoupled-chain
system, the appropriate sense of ``relevant'' is that of the decoupled
1+1-dimensional critical theories. We find
\begin{equation}
\delta V_\times = \sum_{n,m} a_1 (-1)^{n+m}
    \epsilon_{h,m}(na) \, \epsilon_{v,n}(ma)
\label{delta-V-times}
\end{equation}
and
\begin{eqnarray}
\delta V_\parallel &=&
\sum_{n,m,l}\sum_{\nu=h,v}
\Big[ a_2(l) \vec{N}_{\nu,m}(na) \cdot \vec{N}_{\nu,m+l}(na)
\nonumber\\
&&\qquad\qquad{}
    + a_3(l) \epsilon_{\nu,m}(na)\, \epsilon_{\nu,m+l}(na)
\Big].
\quad
\label{delta-V-parallel}
\end{eqnarray}

We note that in Eq.~(\ref{delta-V-parallel}), an interaction is
possible between parallel chains an arbitrary distance $l$ apart.
 From the point of view of the decoupled-chain fixed point, there is no
notion (or effect in RG rescaling) of ``distance'' between chains, so
all such terms are equally ``relevant'' in this point of view.  One
would expect, however, these terms (i.e. $a_2(l), a_3(l)$) to decay
{\sl in magnitude} with increasing $l$.

It is straightforward to check that these terms and only these terms
satisfy the symmetry requirements of the checkerboard lattice.  First,
the invariance of $\delta V_\parallel$ is easy to establish, as it
involves pairs of operators $\epsilon$ and $\vec{N}$ from like chains
(i.e., horizontal-horizontal or vertical-vertical).  These transform
identically under all operations, and invariance is trivially
shown.

The crossed-chains term, $\delta V_\times$, is more involved.  We
sketch the arguments for its invariance.  Rotation by $\pi/2$ about a
crossing is manifest, as the fields in Eq.~(\ref{delta-V-times}) are
drawn from a single such crossing.  Unit translation along the
$x$-direction makes $\epsilon_h \rightarrow -\epsilon_h$ while
$\epsilon_v$ is obviously not affected.  However, $(-1)^{n+m}$ also
changes its sign, $(-1)^{n+m}\rightarrow (-1)^{n+1+m}$, so that
$T_h(\delta V_\times)=\delta V_\times$.  Reflection with respect to a
site on a horizontal chain $P_{sh}$ preserves $\epsilon_v$ but does
change sign of dimerization on every horizontal chain:
$P_{sh}(\epsilon_h)=-\epsilon_h$. But at the same time $P_{sh}$
interchanges even and odd {\em vertical} chains, i.e.,
$P_{sh}((-1)^{n+m})=-(-1)^{n+m}$.  Thus $P_{sh}(\delta
V_\times)=\delta V_\times$.  Link parity $P_{Lh}$ is simple since
every $\epsilon$ is even under it.  Moreover, since $P_{Lh}$ is
nothing but reflection with respect to, say, the vertical chain number
$n$, the vertical chain with index $n+1$ then transforms into that
with index $n-1$, etc.  Hence, even and odd vertical chains are not
interchanged by $P_{Lh}$, and $(-1)^{n+m} \rightarrow (-1)^{n+m}$,
showing the invariance under this final generator.  Notice that the
staggering factor $(-1)^{n+m}$ plays a very important role in this
consideration -- its presence makes the local interaction of staggered
dimerizations possible.

One could wonder if $\delta V_\times$ could similarly include
a staggered product of magnetizations, $(-1)^{n+m} \vec{N}_h \cdot
\vec{N}_h$, but this is prohibited by the $P_{Lh}$ symmetry.  We note
that microscopically, such a term cannot be generated (see the
following subsection for the mechanics of generation of the allowed
terms) as a consequence of the
identity $J^a(x,\tau)\epsilon(x,\tau')=0$ which follows from the OPE
Eq.~(\ref{ope-Je}). The only symmetry-allowed combination of
$\vec{N}$'s that can show up in $\delta V_\times$ is $(\vec{N}_h
\cdot\vec{N}_v)^2$.  Such a term does arise in the large-$S$
``order-from-disorder'' calculations, see Ref.~\onlinecite{olegt}, but
in the $S=1/2$ microscopic model it has scaling dimension $2$ and is
thus deemed irrelevant. Moreover, one can derive, using abelian
bosonization, the OPE of two $\vec{N}$ fields at the same spatial
point $x$: $N^a(x,\tau) N^b(x,0) \sim i\epsilon^{abc}
{\text{sign}}(\tau) J^c(x,0)$.  This allows one to identify
\cite{affleck-thanks} this biquadratic term with the dimension $2$ scalar
product of two spin currents on crossing chains [that is, the $g_{jj}$
term in Eq.~(\ref{V-ccm-full})], $(\vec{N}_h \cdot\vec{N}_v)^2
\rightarrow \vec{J}_h \cdot \vec{J}_v$.

Observe now that none of the symmetry-respecting terms in $\delta
V_\times$ and $\delta V_\parallel$ are present in the na\"ive
continuum limit of the theory Eq.~(\ref{V-ccm-full}).  Below we show
that second-order perturbation theory in the inter-chain exchange
$J_\times$ generates $\delta V_\times$ with coupling constant $a_1
\sim J_\times^2/J$. Similar arguments show that $a_{2,3} \sim
J_\times^4/J^3$.  This is because in $J_\times^2$ order one generates
terms involving a product of derivatives of $\vec{N}$ fields on parallel
chains, $\sim J_\times^2 \partial_x \vec{N}_h \cdot \partial_x \vec{N}_h$.
Once these are present, one can follow calculations in
Ref.~\onlinecite{oleg-leon} to find that both $a_2$ and $a_3$ terms in
$\delta V_\parallel$ are generated, but this happens only in the next,
$(J_\times^2)^2 = J_\times^4$, order of the perturbation expansion.
Since, as we show below in Sec.~\ref{subsec:mf}, the $\delta V_\times$
contribution is relevant, it is sufficient to keep only the leading
$a_1$ type terms---subleading $a_{2,3}$ ones are too small
($a_{2,3}/a_1 \sim J_\times^2/J^2 \ll 1$) to change the outcome.

\subsection{On the importance of irrelevant terms}
\label{subsec:irrel}
Here we describe microscopic calculation of $\delta V_\times$.  We
begin by expanding the action in powers of $V$ Eq.~(\ref{V-ccm-full}).
This generates a number of terms of which the most important ones
involve products of spin currents and staggered magnetizations from
the same chain (and with the same spatial coordinate---that is, all
fields belong to the same crossing). Thus we pick out $g_{jj} g_{nn}$
and cross-terms from $g_{nj}^2$.  These contributions can be written
in the form
\begin{equation}
\sum_{a,b}\sum_{n,m}(-1)^{n+m}\int d\tau d\tau'
v_\times(x,n;y,m;\tau,\tau')\Big|_{x=na,y=ma},
\label{2nd-order}
\end{equation}
where
\begin{eqnarray}
v_\times \!\!&=&\!\!
g_{nj}^2 \left[
\partial_x N^a_{h,m}(x,\tau) J^b_{h,m}(x,\tau')
J^a_{v,n}(y,\tau)\partial_y N^b_{v,n}(y,\tau')
\right.\nonumber\\
&&\left.{}\!\!
+ J^a_{h,m}(x,\tau)\partial_x N^b_{h,m}(x,\tau')
     \partial_y N^a_{v,n}(y,\tau) J^b_{v,n}(y,\tau')\right]
\nonumber\\
&&{}\!\!
+ g_{jj}g_{nn} J^a_{h,m}(x,\tau)\partial_xN^b_{h,m}(x,\tau')
\nonumber\\&&\qquad\times
    J^a_{v,n}(y,\tau) \partial_y N^b_{v,n}(y,\tau').
\label{V^2-long}
\end{eqnarray}
Now apply the OPE Eq.~(\ref{ope-JN}) to the product of fields from the same
chain and at the same {\em spatial point} $x$.
For example,
\begin{eqnarray}
J^a_R(x,\tau)\partial_x N^b(x,\tau')&=&
\lim_{x'\rightarrow x}\partial_{x'}J^a(x,\tau)N^b(x',\tau')
\nonumber\\
&=&
\frac{-i[i\epsilon^{abc} N^c(x,\tau') - i \delta^{ab}\epsilon(x,\tau')]}
{4\pi v^2 (\tau-\tau' + \alpha\sigma_{\tau-\tau'})^2}.
\nonumber\\&&
\end{eqnarray}
Similarly
\begin{equation}
J^a_L(x,\tau)\partial_x N^b(x,\tau')=
\frac{i[i\epsilon^{abc} N^c(x,\tau') + i \delta^{ab}\epsilon(x,\tau')]}
{4\pi v^2 (\tau-\tau' + \alpha\sigma_{\tau-\tau'})^2}.
\end{equation}
We observe that the OPE of the {\em full} spin current $\vec{J}$ and
$\vec{N}$ at the same spatial point does
not contain staggered magnetization
\begin{eqnarray}
J^a(x,\tau)\partial_x N^b(x,\tau')&=&
[J^a_R(x,\tau) + J^a_L(x,\tau)]\partial_x N^b(x,\tau')
\nonumber\\
&=&{}
-\frac{\delta^{ab} \epsilon(x,\tau')}
{2\pi v^2 (\tau-\tau' + \alpha\sigma_{\tau-\tau'})^2}.
\end{eqnarray}
This is a very important result, with which Eq.~(\ref{2nd-order}) can
be brought into the surprisingly compact form:
\begin{eqnarray}
V_2&=&\sum_{a,b} \delta^{ab}\sum_{n,m}(-1)^{n+m} (2g_{nj}^2 +
g_{jj}g_{nn})\nonumber\\
&&\quad\times \int d\tau d\tau'
\frac{\epsilon_{h,m}(na,\tau') \, \epsilon_{v,n}(ma,\tau')}
{[2\pi v^2 (\tau-\tau' + \alpha\sigma_{\tau-\tau'})^2]^2}.
\end{eqnarray}
The integral involved is obviously convergent
\begin{equation}
\int_{-\infty}^\infty dt \frac{1}{(t + \alpha\sigma_t)^4}
=\frac{2}{3\alpha^3}=\frac{2v^3}{3a^3}.
\end{equation}
Using Eq.~(\ref{couplings}) for the $g$'s involved,
we finally obtain the
following fluctuation-generated
correction to the low-energy effective action
\begin{eqnarray}
\delta S&=& - \frac{1}{2} V_2 \nonumber\\
&=&
-\frac{3J_\times^2 a^3}{\pi^2 v}
\int d\tau \sum_{n,m} (-1)^{n+m}
\epsilon_{h,m}(na,\tau) \, \epsilon_{v,n}(ma,\tau).
\nonumber\\&&
\end{eqnarray}
Denoting
\begin{equation}
g_\epsilon=v \frac{3J_\times^2 a^2}{\pi^2 v^2},
\label{g-epsilon}
\end{equation}
we have the following addition to the inter-chain Hamiltonian $V$,
Eq.~(\ref{V-ccm-full}), to analyze
[this is because $Z=\int e^{-S}$ and $\delta S=\int d\tau \delta V$]
\begin{equation}
\delta V_\times =- g_\epsilon a \sum_{n,m} (-1)^{n+m}
\epsilon_{h,m}(na) \epsilon_{v,n}(ma).
\label{delta-V-epsilon}
\end{equation}
Staggered dimerization $\epsilon$ has scaling dimension $1/2$, which
means that it is as
important for the chain physics as $\vec{N}$ is. In fact, up to
logarithmic corrections, correlation
functions of staggered dimerization and magnetization decay with the
same power law, $x^{-1}$, see (\ref{eq:antiferro},\ref{eq:bonden}). 

This is also clear from the OPE Eqs.~(\ref{ope-JN}) and
(\ref{ope-Je}), which
show that $\vec{N}$ and $\epsilon$ transform into each other under
chiral rotations generated by $\vec{J}_{R/L}$.
Since any pair of horizontal and vertical chains has only one crossing,
Eq.~(\ref{delta-V-epsilon}) is a sum of
{\em local} terms, each of which is marginal (space-time dimension
$=1$, and dimension
of the product $\epsilon_h \epsilon_v$ is $1$ as well). However, we
shall see below that this marginality
is superficial---an infinite number of marginal crossings add up to a
relevant perturbation.

\subsection{Mean-field analysis of the effective inter-chain
interaction Eq.~(\ref{delta-V-epsilon})}
\label{subsec:mf}
   From now on it is safe to omit derivative terms present in
$V$, Eq.~(\ref{V-ccm-full}); their role was to generate,
as described in Sec.~\ref{subsec:irrel}, more relevant
symmetry-allowed inter-chain interactions.
With this in mind, we write down the renormalized version of
Eq.~(\ref{V-ccm-full}),
\begin{eqnarray}
V&=&\sum_{n,m}
\Big[
g_{jj} \vec{J}_{h,m}(na) \cdot \vec{J}_{v,n}(ma)
\nonumber\\
&&\qquad{}
- (-1)^{n+m}g_\epsilon a \,
     \epsilon_{h,m}(na) \, \epsilon_{v,n}(ma)
\Big].
     \label{g_jj+g_e}
\end{eqnarray}
As discussed above, the first term originating from the naive
continuum limit of Eq.~(\ref{lattice-V})
has scaling dimension $2$ while the second term, which is $\delta
V_\times$ generated by high-energy fluctuations,
has scaling dimension $1$.
Thus we are allowed to discard the irrelevant current-current piece
of $V$, Eq.~(\ref{g_jj+g_e}).
As a result, all that remains of the interchain interaction is given by
$\delta V_\times$ [Eq.~(\ref{delta-V-epsilon})], $V \to \delta V_\times$,
which was not present in the naive continuum
limit Eq.~(\ref{V-ccm-full}) at all!
We tackle it, in analogy with analysis of Ref.~\onlinecite{oleg-leon},
by the chain mean-field approximation.
The staggering factor $(-1)^{n+m}$ suggests a {\em staggered} dimer
order on parallel chains.
That is, we assume the pattern
\begin{equation}
\epsilon_{h,m}(x)=(-1)^m \langle \epsilon\rangle, \qquad
\epsilon_{v,n}(y)=(-1)^n \langle \epsilon\rangle,
\end{equation}
where $\langle\epsilon\rangle$ is a mean-field expectation value.
The inter-chain coupling is then decoupled into
a sum of independent single-chain Hamiltonians,
\begin{eqnarray}
\delta V_\times&=&
- \sum_m (-1)^m g_\epsilon a
\langle\epsilon\rangle \sum_n \epsilon_{h,m}(x=na) \nonumber\\
&&
- \sum_n (-1)^n g_\epsilon a \langle\epsilon\rangle
    \sum_m \epsilon_{v,n}(y=ma).
\end{eqnarray}
Look on one of them, say, that of the horizontal chain with index $m$
(which is fixed now).
Now we can take continuum limit
($\sum_n f(na) \rightarrow a^{-1}\int dx f(x)$) and
\begin{equation}
\delta V_\times(m) = -(-1)^m g_\epsilon \langle\epsilon\rangle\int
\epsilon(x) \, dx,
\end{equation}
which can be easily analyzed along the lines of
Ref.~\onlinecite{oleg-leon}.
Using abelian bosonization expression for the staggered dimerization
\begin{equation}
\epsilon(x)=\frac{\lambda}{\pi a}\cos[\sqrt{2\pi} \varphi(x)] ,
\end{equation}
where $\varphi$ is the {\em spin} boson field and $\lambda$ is
a nonuniversal constant of order 1,\cite{shelton-ladder}
we arrive at the
effective single-chain
sine-Gordon action for the $m$th chain
\begin{equation}
S(m)=\int d^2r \Big(\frac{1}{2}(\nabla \varphi)^2
    - G \cos\sqrt{2\pi}\varphi\Big).
\end{equation}
The action $S(m)$ is written in terms of dimensionless coordinates
$\vec{r}=(x/a,v\tau/a)$ and the effective coupling constant
$G=\lambda^2 g_\epsilon \langle \cos\sqrt{2\pi}\varphi\rangle/(\pi^2 v)$.
The self-consistent equation for $\langle
\cos\sqrt{2\pi}\varphi\rangle$ follows from the exact
solution \cite{luk} for the free energy $F_m$ of the sine-Gordon
model $Z_m = \int D\varphi \exp[-S(m)] = \exp(-F_m)$;
\begin{equation}
\langle \cos\sqrt{2\pi}\varphi\rangle=-\frac{dF}{dG}=\frac{d\ln
Z}{dG}=\frac{c_0^2}{3\sqrt{3}} G^{1/3} ,
\end{equation}
where the constant $c_0$ reads
\begin{equation}
c_0 = \frac{2\Gamma(1/6)}{\sqrt{\pi}\,\Gamma(2/3)}
\left(\frac{\pi \Gamma(3/4)}{2\Gamma(1/4)}\right)^{2/3}.
\end{equation}
Simple algebra gives
\begin{equation}
\langle \cos\sqrt{2\pi}\varphi\rangle
=\frac{c_0^3}{3}
\sqrt{\frac{\lambda^2 g_\epsilon}{\pi^2 v}}
=0.265316 \frac{J_\times}{J},
\end{equation}
where we have set $\lambda=1$.
Hence the expectation value of the staggered dimerization is proportional
to $J_\times$,
\begin{equation}
\langle\epsilon\rangle=\frac{\lambda}{\pi a}
\langle \cos\sqrt{2\pi}\varphi\rangle
=0.0844\frac{J_\times}{J a}.
\end{equation}
The spin gap $\Delta$ is given by the mass $m$ of the lightest
breather in the sine-Gordon theory \cite{luk}
\begin{equation}
m=2M \sin(\pi/6)=M ,
\end{equation}
where $M$ and $G$ are related by
\begin{equation}
\frac{G}{2}=\frac{\Gamma(1/4)}{\pi \Gamma(3/4)}
\left(M \frac{\sqrt{\pi}\,\Gamma(2/3)}{2 \Gamma(1/6)}\right)^{3/2} .
\end{equation}
Thus $M=c_0 G^{2/3}$, and finally the spin gap $\Delta$ is found as
\begin{equation}
\Delta=m=M=
\frac{4\lambda^2 c_0^3}{\pi^6 \sqrt{3}}
\left(\frac{J_\times}{J}\right)^2=
0.675688 \left(\frac{J_\times}{J}\right)^2.
\end{equation}

The resulting dimerization pattern is shown in Fig.~\ref{patterns}.
An equivalent configuration is obtained by a global shift of crosses
by one lattice spacing along either the $x$ or $y$ direction.  It is
worth pointing out that exactly such inter-dimer correlations
-- crossed-dimer ones -- have been observed in the exact diagonalization
study of finite CCM clusters, see Table II and Fig.~5 in
Ref.~\onlinecite{sindzingre}.

\section{The planar pyrochlore: plaquette VBS and its instabilities}
\label{planar-pyro}

\begin{figure}
\begin{center}
\begin{picture}(240,220)(0,0)
\unitlength=0.9pt
\thicklines
\multiput(20,10)(40,0){6}{\line(0,1){200}}
\multiput(20,10)(0,40){6}{\line(1,0){200}}
\put(20,30){\line(1,1){180}}
\put(20,70){\line(1,1){140}}
\put(20,110){\line(1,1){100}}
\put(20,150){\line(1,1){60}}
\put(20,190){\line(1,1){20}}
\put(40,10){\line(1,1){180}}
\put(80,10){\line(1,1){140}}
\put(120,10){\line(1,1){100}}
\put(160,10){\line(1,1){60}}
\put(200,10){\line(1,1){20}}
\put(20,30){\line(1,-1){20}}
\put(20,70){\line(1,-1){60}}
\put(20,110){\line(1,-1){100}}
\put(20,150){\line(1,-1){140}}
\put(20,190){\line(1,-1){180}}
\put(40,210){\line(1,-1){180}}
\put(80,210){\line(1,-1){140}}
\put(120,210){\line(1,-1){100}}
\put(160,210){\line(1,-1){60}}
\put(200,210){\line(1,-1){20}}
\multiput(20,30)(40,0){6}{\multiput(0,0)(0,40){5}{\circle*{7}}}
\multiput(40,10)(0,40){6}{\multiput(0,0)(40,0){5}{\circle*{7}}}
\multiput(40,30)(80,0){3}{\multiput(0,0)(0,80){3}
    {\textcolor{blue}{\oval(25,25)}}}
\multiput(80,70)(80,0){2}{\multiput(0,0)(0,80){2}
    {\textcolor{blue}{\oval(25,25)}}}
\put(85,107){\textcolor{red}{$\vec{S}_1$}}
\put(115,76){\textcolor{red}{$\vec{S}_2$}}
\put(145,107){\textcolor{red}{$\vec{S}_3$}}
\put(115,137){\textcolor{red}{$\vec{S}_4$}}
\textcolor{red}{
\put(118,100){0}
\multiput(40,30)(40,40){5}{\circle*{2}}
\multiput(40,190)(40,-40){5}{\circle*{2}}
\put(15,5){\vector(1,1){212}}
\put(230,216){\large$j$}
\put(158,140){1}
\put(225,5){\vector(-1,1){212}}
\put(4,215){\large$k$}
\put(78,140){1}
}
\end{picture}
\end{center}
\caption{(Color online) Quadrumerized checkerboard lattice with
  coordinates $(j,k)$ shown.  The plaquettes with blue circles are
  quadrumerized.  Each unit cell contains 4 spins.}
\label{fig:quadrumerized}
\end{figure}
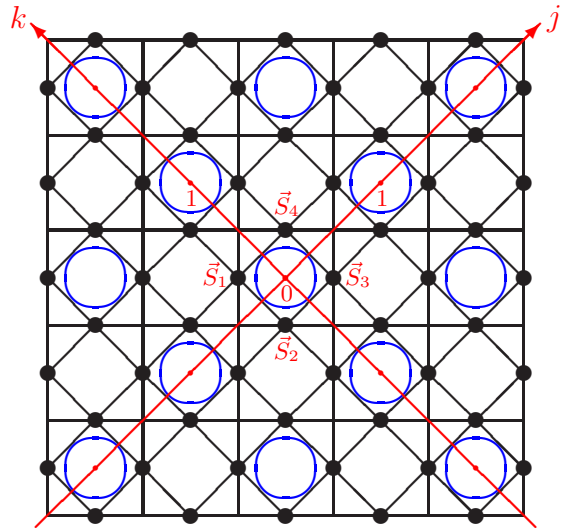

In the preceding sections we focused on the quasi-1D limit
$J_\times\ll J$, and established the existence of the spontaneous
long-range order of the crossed dimer configuration (Fig.~\ref{patterns}).
In this section we will explore a different region in the parameter space,
where the nearest-neighbor coupling $J_\times$
and next-nearest-neighbor exchange coupling $J$ are nearly equal.
Earlier numerical studies using exact diagonalization\cite{fouet} and
strong-coupling expansion techniques\cite{brenig,altman,brenig2}
showed that the ground state at $J=J_\times$ is a valence bond crystal
with long-range quadrumer order, shown in
Fig.~\ref{fig:quadrumerized}.  Here we review a simple theoretical
account of this plaquette VBS (P-VBS) state using the quadrumer-boson
approximation,\cite{starykh96,zhitomirsky,lauchli} and examine its
instabilities to other orders.  This analysis, together with the
results in the preceding sections, will serve as a basis for our
discussion on the global phase diagram of the CCM in the following
section.  Our simple approach presented here is meant to give a
qualitative picture; more quantitatively reliable numerical results
can be obtained, for example, by series expansion, as developed in
Refs.~\onlinecite{brenig} and \onlinecite{brenig2}.

In the following analysis it is more convenient to use
a new coordinate system labelled by $(j,k)$, rotated by $\pi/4$
from the $x$ and $y$ axes; see Fig.~\ref{fig:quadrumerized}.
The quadrumerized valence bond crystal breaks lattice translation
symmetry, and each quadrumerized plaquette centered at $(j,k)$ has
four spins, $\vec{S}_l$ ($l=1,2,3,4$).
 From the outset we assume the breaking of translation symmetry and
begin with the Hamiltonian for a single quadrumerized plaquette,
\begin{eqnarray}
H_p&=&
J_\times\left(
\vec{S}_1\cdot\vec{S}_2+\vec{S}_2\cdot\vec{S}_3
+\vec{S}_3\cdot\vec{S}_4+\vec{S}_4\cdot\vec{S}_1
\right)
\nonumber\\
&=&
\frac{J_\times}{2}\left[
\left(\vec{S}_1+\vec{S}_2+\vec{S}_3+\vec{S}_4\right)^2
-\left(\vec{S}_1+\vec{S}_3\right)^2
\right.\nonumber\\
&&\left.\qquad{}
-\left(\vec{S}_2+\vec{S}_4\right)^2
\right].
\label{H_p}
\end{eqnarray}
The lowest-energy state of $H_p$ is a spin singlet with energy
$-2J_\times$, which can be written as
\begin{eqnarray}
s^\dagger|0\rangle
&=&
\frac{1}{2\sqrt3}\bigl(
|\!\uparrow\uparrow\downarrow\downarrow\,\rangle
+|\!\downarrow\downarrow\uparrow\uparrow\,\rangle
+|\!\uparrow\downarrow\downarrow\uparrow\,\rangle
+|\!\downarrow\uparrow\uparrow\downarrow\,\rangle
\nonumber\\
&&\qquad{}
-2|\!\uparrow\downarrow\uparrow\downarrow\,\rangle
-2|\!\downarrow\uparrow\downarrow\uparrow\,\rangle
\bigr),
\label{eq:singlet}
\end{eqnarray}
where $|\sigma_1 \sigma_2 \sigma_3 \sigma_4\rangle$ denotes the state
with $S^z_l=\sigma_l$.
The first excited states are a triplet with energy $-J_\times$,
\begin{subequations}
\label{eq:triplet}
\begin{eqnarray}
t^\dagger_+|0\rangle\!\!&=&\!\!
\frac{1}{2}\bigl(
|\!\uparrow\uparrow\uparrow\downarrow\,\rangle
+|\!\uparrow\downarrow\uparrow\uparrow\,\rangle
-|\!\uparrow\uparrow\downarrow\uparrow\,\rangle
-|\!\downarrow\uparrow\uparrow\uparrow\,\rangle
\bigr),
\\
t^\dagger_z|0\rangle\!\!&=&\!\!
\frac{1}{\sqrt2}\bigl(
|\!\uparrow\downarrow\uparrow\downarrow\,\rangle
-|\!\downarrow\uparrow\downarrow\uparrow\,\rangle
\bigr),
\\
t^\dagger_-|0\rangle\!\!&=&\!\!
\frac{1}{2}\bigl(
|\!\downarrow\downarrow\downarrow\uparrow\,\rangle
+|\!\downarrow\uparrow\downarrow\downarrow\,\rangle
-|\!\downarrow\downarrow\uparrow\downarrow\,\rangle
-|\!\uparrow\downarrow\downarrow\downarrow\,\rangle
\bigr).
\qquad
\end{eqnarray}
\end{subequations}
The operators $s^\dagger$, $t^\dagger_\pm$, and $t^\dagger_z$
can be thought of as creation operators of hard-core bosons.

As mentioned above, the ground state of the CCM is known to be a
gapped P-VBS state at the planar pyrochlore point $J=J_\times$.
As long as $J\approx J_\times$, we may thus expect that
a good approximation to the ground state should be obtained
by direct product of the singlet states, Eq.~(\ref{eq:singlet}),
weakly hybridized with the triplets, Eqs.~(\ref{eq:triplet}).
Motivated by this observation, we employ the quadrumer-boson
approximation\cite{starykh96,zhitomirsky,lauchli}
in which we keep only the
low-lying four states, singlet and triplet,
in each quadrumerized plaquette,
and discard the other higher-energy states.
Now the boson operators are subject to the constraint
\begin{equation}
s^\dagger s + t^\dagger_+t^{}_+ + t^\dagger_zt^{}_z
+t^\dagger_-t^{}_-
=1.
\end{equation}
The plaquette Hamiltonian is then written as
\begin{equation}
H_p=
-2J_\times
+J_\times\left(
t^\dagger_+t^{}_++t^\dagger_zt^{}_z+t^\dagger_-t^{}_-
\right).
\end{equation}
The spins $\vec{S}_l$ can also be written in terms of the hard-core
boson operators.
The representations are found from matrix elements of the spin
operators with the four states.
After some algebra we find
\begin{subequations}
\begin{eqnarray}
S^z_l&=&
\frac{1}{4}\left(t^\dagger_+t^{}_+-t^\dagger_-t^{}_-\right)
+\frac{(-1)^l}{\sqrt6}\left(t^\dagger_zs+s^\dagger t^{}_z\right),
\\
S^+_l&=&
\frac{1}{\sqrt8}\left(t^\dagger_+t^{}_z-t^\dagger_zt^{}_-\right)
-\frac{(-1)^l}{\sqrt3}\left(t^\dagger_+s+s^\dagger t^{}_-\right),
\qquad
\\
S^-_l&=&
\frac{1}{\sqrt8}\left(t^\dagger_zt^{}_+-t^\dagger_-t^{}_z\right)
-\frac{(-1)^l}{\sqrt3}\left(t^\dagger_-s+s^\dagger t^{}_+\right),
\end{eqnarray}
\end{subequations}
where $l=1,2,3,4$.
Assuming that the density of triplets is low in the P-VBS state,
we keep only terms linear in $t_\mu$ and
set $s=s^\dagger=1$;
\begin{equation}
S^a_l=
\frac{(-1)^l}{\sqrt6}\left(t^\dagger_a+t^{}_a\right),
\label{eq:S_i}
\end{equation}
where $a=x,y,z$ and we have introduced
\begin{equation}
t^{}_x=-\frac{1}{\sqrt2}(t^{}_+ + t^{}_-),
\qquad
t^{}_y=\frac{1}{\sqrt2 i}(t^{}_+ - t^{}_-).
\end{equation}

With the coordinate system $(j,k)$ in Fig.~\ref{fig:quadrumerized}
the total Hamiltonian of the checkerboard antiferromagnet reads
\begin{eqnarray}
H\!\!&=&\!\!
\sum_{j,k}H_p(j,k)
\nonumber\\
&&{}
+J_\times\sum_{j,k}\left(
\vec{S}_{j,k,1}\cdot\vec{S}_{j-1,k,4}
+\vec{S}_{j,k,2}\cdot\vec{S}_{j,k-1,1}
\right.\nonumber\\
&&\left.\qquad\qquad{}
+\vec{S}_{j,k,3}\cdot\vec{S}_{j+1,k,2}
+\vec{S}_{j,k,4}\cdot\vec{S}_{j,k+1,3}
\right)
\nonumber\\
&&{}
+J\sum_{j,k}\left(
\vec{S}_{j,k,1}\cdot\vec{S}_{j,k+1,3}
+\vec{S}_{j,k,4}\cdot\vec{S}_{j,k+1,2}
\right.\nonumber\\
&&\left.\qquad\qquad{}
+\vec{S}_{j,k,3}\cdot\vec{S}_{j+1,k,1}
+\vec{S}_{j,k,4}\cdot\vec{S}_{j+1,k,2}
\right),
\nonumber\\&&
\label{eq:H3}
\end{eqnarray}
where $\vec{S}_{j,k,l}$ is the $l$th spin $\vec{S}_l$ in
the quadrumerized plaquette centered at $(j,k)$.
With the approximation Eq.~(\ref{eq:S_i}) the Hamiltonian becomes
\begin{equation}
H=
\sum_{\vec{p}}
\left(
-\frac{7}{2}J_\times
+\frac{1}{2}\sum_{a=x,y,z}
\Psi^\dagger_a(\vec{p})\mathcal{H}(\vec{p})\Psi^{}_a(\vec{p})
\right),
\label{eq:HPsi}
\end{equation}
where we have introduced the triplet boson field,
\begin{equation}
\Psi_a(\vec{p})=
\begin{pmatrix}
\tilde{t}^{}_a(\vec{p}) \\ \\
\tilde{t}^\dagger_a(-\vec{p})
\end{pmatrix},
\end{equation}
with the momentum $\vec{p}=(p_1,p_2)$ and the Fourier transform
\begin{equation}
\tilde{t}_a(\vec{p})=\frac{1}{\sqrt{\cal{N}_\circ}}\sum_{j,k}
e^{-i(jp_1+kp_2)}t_a(j,k) ,
\end{equation}
where $\cal{N}_\circ$ is the number of quadrumerized plaquettes.
The Hamiltonian matrix is given by
\begin{equation}
\mathcal{H}(\vec{p})=
\begin{pmatrix}
J_\times+\varepsilon(\vec{p}) &
\varepsilon(\vec{p}) \\ \\
\varepsilon(\vec{p}) &
J_\times+\varepsilon(\vec{p})
\end{pmatrix},
\end{equation}
where
\begin{equation}
\varepsilon(\vec{p})=\frac{2}{3}(J-J_\times)(\cos p_1 + \cos p_2).
\end{equation}
With the Bogoliubov transformation,
\begin{equation}
\begin{pmatrix}
\tilde{t}^{}_a(\vec{p}) \\ \tilde{t}^\dagger_a(-\vec{p})
\end{pmatrix}=
\begin{pmatrix}
\cosh\theta_{\vec{p}} & \sinh\theta_{\vec{p}} \\
\sinh\theta_{\vec{p}} & \cosh\theta_{\vec{p}}
\end{pmatrix}
\begin{pmatrix}
b^{}_a(\vec{p}) \\ b^\dagger_a(-\vec{p})
\end{pmatrix},
\end{equation}
where
\begin{equation}
\exp(-4\theta_{\vec{p}})=1+\frac{2\varepsilon(\vec{p})}{J_\times},
\end{equation}
the Hamiltonian [Eq.~(\ref{eq:HPsi})] is diagonalized,
\begin{equation}
H
=\sum_{\vec{p}}
\left(
-\frac{7}{2}J_\times
+\frac{3}{2}E(\vec{p})
+E(\vec{p})\sum_a
b^\dagger_a(\vec{p})b^{}_a(\vec{p})
\right).
\end{equation}
The energy dispersion of the triplet eigen mode $b_a(\vec{p})$
is given by
\begin{equation}
E(\vec{p})=
\left[J_\times
\left(J_\times+\frac{4}{3}(J-J_\times)(\cos p_1+\cos p_2)\right)
\right]^{1/2}.
\label{triplet_dispersion}
\end{equation}

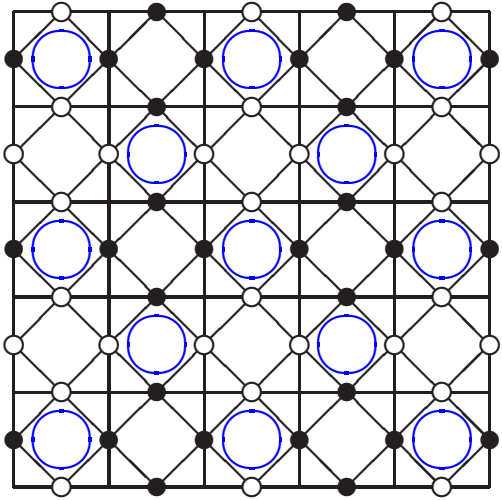
\begin{figure}
\begin{center}
\begin{picture}(240,213)(0,5)
\unitlength=0.9pt
\thicklines
\multiput(20,10)(40,0){6}{\line(0,1){200}}
\multiput(20,10)(0,40){6}{\line(1,0){200}}
\put(20,30){\line(1,1){180}}
\put(20,70){\line(1,1){140}}
\put(20,110){\line(1,1){100}}
\put(20,150){\line(1,1){60}}
\put(20,190){\line(1,1){20}}
\put(40,10){\line(1,1){180}}
\put(80,10){\line(1,1){140}}
\put(120,10){\line(1,1){100}}
\put(160,10){\line(1,1){60}}
\put(200,10){\line(1,1){20}}
\put(20,30){\line(1,-1){20}}
\put(20,70){\line(1,-1){60}}
\put(20,110){\line(1,-1){100}}
\put(20,150){\line(1,-1){140}}
\put(20,190){\line(1,-1){180}}
\put(40,210){\line(1,-1){180}}
\put(80,210){\line(1,-1){140}}
\put(120,210){\line(1,-1){100}}
\put(160,210){\line(1,-1){60}}
\put(200,210){\line(1,-1){20}}
\multiput(20,70)(40,0){6}{\multiput(0,0)(0,80){2}
    {\textcolor{white}{\circle*{8}}}}
\multiput(20,70)(40,0){6}{\multiput(0,0)(0,80){2}{\circle{8}}}
\multiput(20,30)(40,0){6}{\multiput(0,0)(0,80){3}{\circle*{8}}}
\multiput(80,10)(0,40){6}{\multiput(0,0)(80,0){2}{\circle*{8}}}
\multiput(40,10)(0,40){6}{\multiput(0,0)(80,0){3}
    {\textcolor{white}{\circle*{8}}}}
\multiput(40,10)(0,40){6}{\multiput(0,0)(80,0){3}{\circle{8}}}
\multiput(40,30)(80,0){3}{\multiput(0,0)(0,80){3}
    {\textcolor{blue}{\oval(24,24)}}}
\multiput(80,70)(80,0){2}{\multiput(0,0)(0,80){2}
    {\textcolor{blue}{\oval(24,24)}}}
\end{picture}
\end{center}
\caption{(Color online) The magnetically ordered state due to condensation
  of bosons $b_a(\pi,\pi)$.  The solid circles represent, say, up
  spins, and the open circles represent down spins.  The quadrumerized
  plaquettes in the nearby P-VBS phase are indicated by blue circles.
}
\label{fig:order}
\end{figure}

The simple quadrumer-boson approximation described above reproduces
the basic feature of the previous numerical
studies:\cite{fouet,brenig,altman,brenig2}
at $J\approx J_\times$ the P-VBS state is a stable
singlet ground state with a gap to excited states.

The softening of the triplet mode Eq.~(\ref{triplet_dispersion}) tells us
potential instabilities that the P-VBS state may have (see also
Ref.~\onlinecite{brenig2}).  First, from
$E(0)=[J_\times(8J-5J_\times)/3]^{1/2}$ we see that it becomes
unstable at $J_\times\to 8J/5$, when the bosons condense at
$\vec{p}=0$.  The resulting state has the long-range N\'eel order,
like the one realized in the square-lattice limit $J_\times\gg J$.
This can be easily seen from Eq.~(\ref{eq:S_i}), in which we may
suppose that $t^\dagger_a + t^{}_a$ is a nonvanishing $c$-number upon
condensation of bosons.  In the present approximation, the transition
from the plaquette singlet to the N\'eel-ordered state occurs at
$(J/J_\times)_{c1}=5/8$, which should be compared with the estimate
$(J/J_\times)_{c1}=0.79\sim0.81$ from a sophisticated strong-coupling
expansion.\cite{brenig2} Second, the P-VBS state becomes unstable as
$J/J_\times\to(J/J_\times)_{c2}=11/8$, at which bosons with momentum
$\vec{p}=(\pi,\pi)$ condense.  Upon bose condensation the spins will
have a magnetic long-range order, shown in Fig.~\ref{fig:order}, with
the spin configuration $\uparrow\uparrow\downarrow\downarrow
\uparrow\uparrow\downarrow\downarrow$ along the diagonal directions
and the N\'eel order along the horizontal and vertical chain
directions [as can be seen by replacing $t_a^\dagger + t_a \to
(-1)^{j+k}$ in Eq.~(\ref{eq:S_i})].  In fact, this is one of the
magnetically ordered states which are shown to be stable at $J\gg
J_\times$ in the $1/S$ expansion; see Fig.~2(b) and Fig.~6 in
Ref.~\onlinecite{olegt}.  The softening of the triplet mode at
$\vec{p}=(\pi,\pi)$ is also found in the strong-coupling expansion in
Ref.~\onlinecite{brenig2} with the estimated critical coupling
$(J/J_\times)_{c2}=1.06\sim1.13$.

Finally, we comment on two defects in our approach.  One is that the
lattice translation symmetry is explicitly broken from the outset and
cannot be restored within the theory.  This means that the N\'eel
ordered state at $J_\times>5J/8$ inevitably has the P-VBS order as
well -- i.e., it is a coexistence phase.  This should probably be viewed
as an artifact of the approach -- see Sec.~\ref{global} for discussion
of this portion of the phase diagram.  The other defect is that we
have ignored interactions among the triplet bosons (as well as
projected out the other higher-energy states -- singlet, triplet, and
quintuplet -- on each quadrumerized plaquette).  In the crudest
approximation we adopted, the first excited states are a triplet
excitation $S=1$ with no energy dispersion at $J_\times=J$.  The
numerical studies\cite{fouet,brenig,altman,brenig2} showed, however,
that at $J=J_\times$ the lowest-excited states are in the spin singlet
sector (see, for example, Sec.\ V in Ref.~\onlinecite{fouet}).  The
cluster-based calculations\cite{brenig,altman} indicate that these
singlet excitations are bound states of two $S=1$ excitations.  To
describe correctly the singlet excitations in terms of the quadrumer
bosons, one needs to go beyond the linear approximation Eq.~(\ref{eq:S_i})
and include interactions among the bosons.  We do not try to do this
here, but only point out that the {\em dispersionless} triplet mode is
naturally susceptible to forming a bound state.  Away from the planar
pyrochlore point $J_\times=J$, not much is known about the low-energy
excitations in the $S=0$ sector; it is not well understood how the
singlet energy levels in the spin gap change as a function of
$J_\times/J$.\cite{brenig-discussion}

\section{Global phase diagram of the CCM}
\label{global}

In this section we discuss the global zero-temperature phase diagram
of the CCM as the control parameter $J_\times/J$ is increased from $0$
to $\infty$.  Our analysis relies on three well-established facts.
First, the ground state at $J_\times\gg J$ obviously has the N\'eel
order and is smoothly connected to the N\'eel ordered state of the
antiferromagnetic Heisenberg model on the square lattice.  Second, by
now there is convincing numerical \cite{fouet,brenig,altman,brenig2}
and analytical \cite{sachdev,moessner,toronto} evidence for the P-VBS 
state at and around
$J=J_\times$.  Finally, as shown in Sec.~\ref{sec:solution}, the
ground state is spontaneously dimerized in the quasi-one-dimensional
$J_\times \ll J$ limit as well, where long-range crossed-dimer order
(Fig.~\ref{patterns}) sets in.

The first question we address here is how the two dimerized phases,
plaquette VBS (P-VBS) and crossed-dimer VBS (CD-VBS), are connected.
We propose the two complementary scenarios in subsections
\ref{subsec:1order} and \ref{subsec:ordered} below.
The nature of the transition between the P-VBS and the N\'eel states
is discussed in Sec.~\ref{subsec:p-vbs-neel}.

\begin{figure}
\center
\includegraphics[width=\columnwidth]{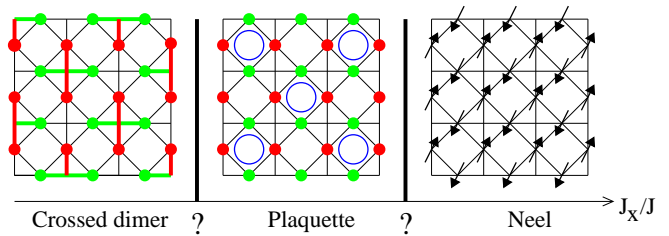}
\caption{(Color online) Global phase diagram of the CCM model in scenario I of
    Sec.~\ref{subsec:1order}. Thick vertical lines with question
    marks indicate that the corresponding transition is, according to
    Landau theory reasoning, either {\sl first-order} or occurs via an
    intermediate {\sl co-existence} phase.}
\label{fig:global-1}
\end{figure}

\subsection{Scenario I: direct transition between the crossed-dimer and
plaquette VBS}
\label{subsec:1order}
The `minimal' assumption is that the two quantum-disordered valence
bond phases connect at some critical value $J_\times/J < 1$. The
validity of this assumption can only be verified by the {\sl exact}
calculation of the ground state energies of two dimerized phases in
the whole interval $0 < J_\times/J < 1$ of interest.  Such a
calculation is obviously beyond our analytic approximations suited for
$J_\times/J \to 0$ (CD-VBS, Sec.~\ref{sec:solution}) and $J_\times/J
\to 1$ (P-VBS, Sec.~\ref{planar-pyro}) limits.  Instead, we take a
phenomenological point of view here and assume that the ground state
energies are such that a {\sl direct} transition between the CD-VBS
and P-VBS phases is possible. At least partial support to this point
of view is provided by the exact diagonalization study of Sindzingre
{\it et al.}\cite{sindzingre} which seems to indicate a single change
in the ground state around $J_\times/J \approx 0.8$.

The question then is if this transition between these two phases can
be continuous.  In general, this question is difficult to answer.
Most formally, the renormalization group theory of continuous critical
phenomena sets only some rather weak constraints on the existence
of a continuous phase transition between any two phases ``A'' and
``B''.  In particular, it requires the existence of an abstract
scale-invariant fixed point (critical field theory) with a single
``relevant'' symmetry-allowed operator in its spectrum, such that a
positive/negative coefficient of this operator in the action takes the
system into phase A/B.  The critical fixed point theory itself must
clearly have {\sl higher} symmetry than either phase A or B, but no a
priori restriction is placed on the relation of the symmetries of
phase A to those of phase B.

A conventional -- and more stringent -- ``criterion'' for the existence of
a continuous transition is based on the specific realization of the
critical theory provided by a Landau-Ginzburg-Wilson (LGW) action
written in terms of order parameters.  More physically, LGW theory
permits a continuous transition by the ``condensation'' of some ``soft
mode'' of phase A, which transforms non-trivially under the symmetry
group of A.  The condensation of this soft-mode order parameter then
leads to a lowering of symmetry (since by assumption the condensation
breaks some symmetries that it transforms under) in phase B.  A
necessary criterion for a LGW-allowed continuous phase transition is
thus that the symmetry group of one phase (B in the example) is a
sub-group of the other (phase A).\cite{landau-statmech}
Further restrictions are implied by
detailed consideration of the LGW expansion (e.g., presence of cubic
terms, etc.), as is standard.\cite{landau-statmech}

Recent work on related but distinct quantum phase transitions has
provided explicit theoretical examples of non-LGW critical
theories,\cite{dcp} demonstrating that the violation of this
conventional ``criterion'' is more than a formal possibility.
Unfortunately, there is at present no general prescription to supplant
the LGW criterion, so we are left in the uncomfortable position of
being unable to solidly argue for or against the possibility of a
continuous quantum critical point (QCP).
Instead, we will content ourselves here with the LGW analysis.

It is straightforward to conclude that a continuous transition between
the CD-VBS and P-VBS states is prohibited by the LGW criterion.  This
can be seen by the two lattice reflections ${\cal R}_{1,2}$, which map
the crossed-chains lattice onto itself, i.e., symmetries of the
Hamiltonian. Here ${\cal R}_1$ is the reflection with respect to a
horizontal chain [it corresponds to a link parity operation $P_L$
Eq.~(\ref{P_L}) on all vertical chains], and ${\cal R}_2$ is the
reflection with respect to a horizontal line passing through the
centers of empty plaquettes [this is a site parity $P_s$ Eq.~(\ref{P_s})
from the point of view of vertical chains]; similar reflections with
respect to vertical lines are accounted for by the $\pi/2$ rotational
symmetry (about chain crossings) of the lattice.

Both phases are two-fold degenerate, as can been seen from
Fig.~\ref{fig:global-1} (and hence can be described by Ising order
parameters).  Their symmetries are distinct.  In particular, note
first that ${\cal R}_1$ is a symmetry of the CD-VBS phase, but not the
P-VBS (it interchanges the two  P-VBS ground states).  Thus the
symmetry group of the CD-VBS phase is not a subgroup of that of the
P-VBS phase.  Second, ${\cal R}_2$ is a symmetry of the P-VBS phase,
but not the CD-VBS phase.  Thus the symmetry group of the P-VBS phase
is not a subgroup of the CD-VBS state.  Since neither symmetry group
is a subgroup of the other, a continuous LGW transition between the
two states is not possible, as promised.

The simplest alternative is a first order transition between the two
phases, which is always possible, and may perhaps be likely.  Another
possibility, is that, between the two states, there is a finite range
of coexistence of P-VBS and CD-VBS order. Such a coexistence phase can
have continuous LGW (Ising) transitions to both the P-VBS and CD-VBS
states.  The latter scenario is only one of a multitude of
conceptually possible phase structures, for which we have no physical
motivation.  We indicate this uncertainty by the question mark in
Fig.~\ref{fig:global-1}.

\subsection{Scenario II: CD to P-VBS via an intermediate ordered phase}
\label{subsec:ordered}

\begin{figure}
\center
\includegraphics[width=0.95\columnwidth]{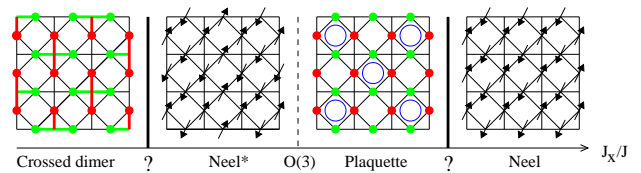}
\caption{(Color online) Global phase diagram of the CCM model in scenario II of
  Sec.~\ref{subsec:ordered}. The continuous $O(3)$ transition between
  the N\'eel$^*$ and P-VBS phases is indicated by a dashed vertical
  line.  Other notations are as in Fig.~\ref{fig:global-1}.}
\label{fig:global-2}
\end{figure}

A quite different scenario is suggested by the quadrumer-boson
approximation of Sec.~\ref{planar-pyro}: an intermediate {\sl
    magnetically ordered} phase between the CD-VBS and P-VBS states.  It
was found there that the P-VBS state becomes unstable at
$(J_\times/J)_{c2}=8/11$ as the $J_\times/J$ ratio is reduced below
the planar pyrochlore value of $1$.  The resulting state, depicted in
Fig.~\ref{fig:order} and Fig.~\ref{fig:global-2}, possesses long-range
magnetic order. We denote it as the N\'eel$^*$ state in the following.
This magnetically ordered state was previously found in the large-$S$
approach \cite{olegt}, where it arises as a result of a quantum
``order-from-disorder'' selection among the large family of degenerate
(at $S=\infty$) collinear ordered states. The fact that it also
appears as a result of the triplet softening of the $S=1/2$ P-VBS
phase of Sec.~\ref{planar-pyro} gives a strong independent argument in
favor of its stability. Of course, as in the previous scenario,
the ultimate fate of the N\'eel$^*$ phase is to be decided by
accurate numerical investigations, and here we just assume that this
ordered phase is indeed the ground state in a finite $J_\times/J$
window within the $(0,1)$ interval.

The transition between the P-VBS and the N\'eel$^*$ phases manifests
itself as a triplet condensation transition within the consideration
of Sec.~\ref{planar-pyro}.  Since such a ``soft-mode'' transition is
the general physical interpretation of LGW theory, it should not come
as a surprise that the N\'eel$^*$ to P-VBS transition satisfies the
LGW criterion for a continuous critical point.  In particular, the
symmetry group of the N\'eel$^*$ state is generated by: (1) spin
rotations about the ordering axis, (2) translations along a diagonal
[$(1,1)$ or $(1,-1)$ in the coordinate system of
Fig.~\ref{fig:global-2}] composed with a time-reversal which inverts
the spins $\vec{S}_r \rightarrow -\vec{S}_r$, (3) reflections ${\cal
    R}_2$, and (4) $\pi/2$ rotations about the center of a plaquette
containing four parallel spins.  Plainly, from Fig.~\ref{fig:global-2},
every one of these operations is also a symmetry of the P-VBS state;
hence the symmetry group of the N\'eel$^*$ state is a subgroup of that
of the P-VBS state.  Moreover, the triplet condensation amplitude can
be identified with the N\'eel$^*$ order parameter: an $O(3)$ vector
specifying the direction of spin orientation at some reference site in
Fig.~\ref{fig:global-2}.  Indeed, an LGW expansion could be developed
in this order parameter, but we content ourselves with the expectation
that the P-VBS to N\'eel$^*$ transition is likely in the continuous
$O(3)$ universality class.

Clearly, the N\'eel$^*$ phase cannot survive down to the
$J_\times/J=0$ point which describes decoupled $S=1/2$ spin chains,
as quantum spin fluctuations destroy antiferromagnetic
N\'eel long-range order in individual chains.\cite{olegt}  The
calculations of Sec.~\ref{sec:solution} demonstrate that at small
$J_\times/J$, the fluctuating dimerization field $\epsilon$ takes
over the (quasi-classical) spin fluctuations and drives the chains
into the CD-VBS phase.

We can again apply the LGW criterion to ask whether a continuous
CD-VBS to N\'eel$^*$ transition is possible.  Clearly, the symmetry
group of the CD-VBS state cannot be a subgroup of the N\'eel$^*$
phase, since the former is spin-rotationally invariant.  However, the
N\'eel$^*$ phase is invariant under ${\cal R}_2$, which, as we saw in
the previous subsection, is not a symmetry of the CD-VBS phase.  Thus,
the symmetry group of the N\'eel$^*$ state is not a subgroup of that
of the CD-VBS phase, and a continuous LGW transition between these two
phases is prohibited.

The transition then is likely either a {\sl first-order} one or
proceeds via an intermediate ordered and bond-modulated {\sl
    coexistence} phase. Such a phase is easy to imagine --- start with the
N\'eel$^*$ state and modulate slightly spin exchange couplings along
horizontal and vertical chains so that ``strong'' bonds repeat the
pattern of dimers in the CD-VBS phase. This state clearly breaks
${\cal R}_2$, but does preserve the long-range magnetic order (for
sufficiently weak modulation) of the N\'eel$^*$ phase. The transition
from the CD-VBS to such a ``modulated'' N\'eel$^*$ one is then
continuous $O(3)$ spin-symmetry breaking, while at some higher
$J_\times/J$ the bond modulation goes away via a continuous $Z_2$
transition and one obtains the pure N\'eel$^*$ phase. Which of the two
possibilities, first-order or coexistence, is realized cannot be
decided within our analytical approach, and for this reason the
transition between CD-VBS and N\'eel$^*$ phases is marked by a
question mark in Fig.~\ref{fig:global-2}.

\subsection{Plaquette VBS to N\'eel transition}
\label{subsec:p-vbs-neel}

Regardless of the phase structure for $J_\times/J <1$, we expect a
transition at larger $J_\times$ from the P-VBS state to the
conventional N\'eel state.  As is the case with the CD-VBS to
N\'eel$^*$ transition discussed above,
a continuous P-VBS to N\'eel QCP is forbidden within
LGW theory (by similar arguments, which we refrain from giving
explicitly for brevity).  In this particular case, however, an alert
reader may note that the two phases appear to be very similar to those
recently argued to be connected by a continuous but non-LGW continuous
phase transition, deemed a {\sl Deconfined} Quantum Critical Point
(DQCP).\cite{dcp}

The analogy, however, is not complete.  Significantly, the
checkerboard lattice differs in detail from the square lattice
discussed in Ref.~\onlinecite{dcp} in its point symmetry group. In
particular, a $\pi/2$ rotation about a {\sl site} of the lattice, a
symmetry of the square lattice, is {\sl not} a symmetry operation of
the checkerboard lattice.  We believe this symmetry distinction is
sufficient to destabilize the putative DQCP.  It is beyond the scope
of this paper to fully recapitulate the arguments of
Ref.~\onlinecite{dcp}, which would be necessary to explain this
conclusion in a stand-alone fashion.  Instead, we will sketch these
arguments, assuming the reader will refer to Ref.~\onlinecite{dcp} for
further details.

The crucial, indeed defining, property of the DQCP is an emergent
topological conservation law, exactly maintained at the critical fixed
point.  Specifically, ``skyrmion number'' is conserved by the fixed
point theory.  This is not true microscopically at the lattice level,
but is an emergent feature of critical theory, as argued in
Ref.~\onlinecite{dcp}.  A crucial step in that argument is the
remarkable identification (due to Haldane\cite{haldane88})
of the skyrmion creation operator with the columnar/plaquette
VBS order parameter.  These can
be defined through a complex scalar field $\Psi$ (see
Ref.~\onlinecite{dcp}). We emphasize that although $\Psi$ has the
symmetries of the rather ``conventional'' VBS order parameter, it is
not to be viewed as a nearly-free field in the LGW sense, but rather
some ``composite'' operator in the critical field theory.  Under the
$\pi/2$ site rotation above, one finds $\Psi \rightarrow i\Psi$, and
consequently, for the square lattice, the skyrmion creation operator
can appear only in the fourth order in the continuum field theory
action for the square lattice antiferromagnet, i.e., as a perturbation
of the form $S'=\int \! d\tau d^2x \, \lambda_4 {\rm Re}\, \Psi^4$.  A
variety of arguments indicate then, because of the large (fourth)
power of $\Psi$ which appears, $\lambda_4$ is {\sl irrelevant} at the
DQCP.  On the checkerboard lattice, lacking this $\pi/2$ site
rotation, the remaining symmetries of the system are insufficient to
rule out the much more relevant ``quadratic'' term $S'=\int \!  d\tau
d^2x \, \lambda_2 {\rm Re}\, \Psi^2$.  The presence of the non-zero
$\lambda_2$ term is, incidentally, also tied to the only {\sl
    two-fold} degeneracy of the P-VBS state, compared to the {\sl
    four-fold} degeneracy of the columnar and plaquette VBS states on
the square lattice.

A necessary and sufficient condition for the stability of this DQCP on
the checkerboard lattice is thus the irrelevance of $\lambda_2$.
Ultimately, this can be decided only by detailed numerical
calculations of the scaling dimension of the $\Psi^2$ operator.
Unpublished numerical results\cite{motrunichprivate}\ for the
easy-plane deformation of the theory (which is more numerically
tractable) suggest that it is in fact {\sl relevant}.  If this
conclusion is, as we suspect, true for the SU(2) symmetric
model, {\sl the DQCP is not stable on the checkerboard lattice}.  Thus
we are led to conclude that there is no viable candidate theory for a
continuous N\'eel to P-VBS quantum critical point in this model, and
that such a transition is quite unlikely.

The simple harmonic analysis of the previous section predicts another
``soft mode'' transition in the $O(3)$ universality class out of the
quadrumerized VBS state to one with N\'eel order at $J/J_\times<1$.
The resulting magnetically ordered phase is, by its very construction,
a coexistence region with both P-VBS and N\'eel order, that is, with
{\sl less} symmetry than either phase.  This is built into the
quadrumer boson expansion because all excitations are constructed
about a background that explicitly has the reduced symmetry of the
P-VBS state, and there is no mechanism to restore the full point group
symmetries of the checkerboard lattice.  Thus we believe the
alternative possibility of a direct first-order (since the DQCP theory
is unstable) transition from the P-VBS to the true N\'eel state should
not be ruled out as a possibility.  The possible existence of a
coexistence region between VBS and N\'eel orders in various models is
still a subject of some contention.  It has been discussed in great
detail in Ref.~\onlinecite{sachdev-park}. An exact diagonalization study
of the quantum checkerboard antiferromagnet,\cite{sindzingre} has
concluded that, if present at all, the co-existence phase is very
narrow. Clearly more detailed studies of this interesting question are
needed. At present, we can only reiterate that a single continuous
transition is highly unlikely in view of the arguments presented
above.

\section{Back to three dimensions}
\label{sec:3d}

\begin{figure}
\center
\includegraphics[width=0.6\columnwidth]{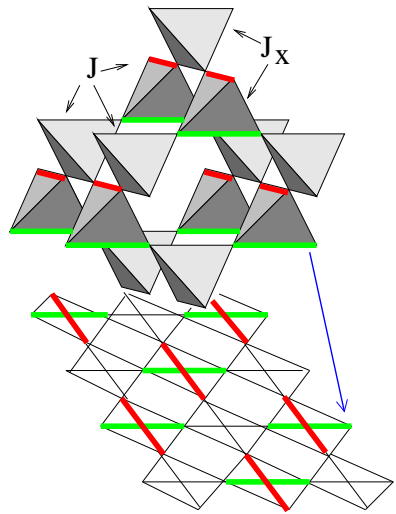}
\caption{(Color online) Three-dimensional dimerization pattern of a generalized
  ``quasi-one-dimensional" pyrochlore antiferromagnet.  Also shown, by
  a long blue arrow, is its ``two-dimensional'' projection, which
  coincides with the crossed-dimer order of Fig.~\ref{patterns}.}
\label{fig:pyro3d}
\end{figure}

Consider now the 3D pyrochlore. Although all bonds of a
tetrahedron are equivalent by symmetry, it makes sense to ask, in
analogy with the 2D lattice that we analyzed in this paper, what would
happen if some bonds were stronger than others. The particular
generalization motivated by the present study of the 2D projected
model (which can be thought of as a ``shadow" of the 3D pyrochlore on 
a 2D plane), involves
a modified model in which two opposite bonds of tetrahedron are strong
($J$) whereas four remaining bonds, connecting the strong ones, are
weak ($J_\times$).  Then, in the limit $J_\times/J \ll 1$, one is back
to the problem of strong spin chains coupled by weak and frustrated
inter-chain $J_\times$. Now, however, chains are arranged in layers: chains
are parallel to each other (oriented along either $x$ or $y$
direction) in each layer, but are orthogonal to those in the layers
right above and below.  That is, chains form a stack of the type
$x-y-x-y...$ along the vertical ($z$) direction.  Chains in one layer
do not interact with each other, $J_\times$ couples orthogonal chains
from neighboring layers. This is just a 3D generalization of 2D
situation analyzed in this paper. It does not introduce any new features,
and hence the answer for the ground state is straightforward -- it is
spontaneously dimerized into the pattern shown in
Fig.~\ref{fig:pyro3d}.

Such a generalization is not unrealistic. It appears that $S=1/2$
pyrochlore material GeCu$_2$O$_4$ has exactly such a
quasi-one-dimensional structure,\cite{gecuo} thanks to a strong
Jahn-Teller elongation of CuO$_6$ octahedra along the crystal $c$
direction.\cite{gecuo} From the high-temperature tail of the uniform
spin susceptibility one can estimate the ratio of (frustrated)
inter-chain exchange $J_\times$ to the intra-chain one $J$ as
$J_\times/J \approx 0.16$.\cite{gecuo} At lower temperatures the
uniform spin susceptibility follows that of the spin chain down to
$T_c = 33$\,K, where a small discontinuity is observed. The specific
heat shows a sharp peak at the same temperature, suggesting a
first-order transition to the magnetically ordered state of yet
unknown structure (as far as we know, Ref.~\onlinecite{gecuo} is the
only experimental study of this interesting material at the moment).
The theory presented in this paper predicts that for a sufficiently
small $J_\times/J$ ratio, the ordered state will be replaced by the
quantum-disordered valence-bond solid shown in Fig.~\ref{fig:pyro3d}.

Another very interesting realization of ``one-dimensionality" in the
3D setting is provided by the
$S=1$ pyrochlore material ZnV$_2$O$_4$.\cite{znvo} There spin chains
are formed below the structural
{\em orbital-ordering} transition at $T_{c1} =50$\,K, as observed in
the recent neutron scattering experiment.\cite{shlee}
This is followed by the second, {\em magnetic}, transition at
$T_{c2}=40$~K. The resulting collinear  magnetic
order can be described
as follows: $S=1$ spins order antiferromagnetically along the strong
(chain) directions in such a way that
spins along weaker ($J_\times$) bonds form a ``4-spin"
pattern,\cite{tsunetsugu}
``up-up-down-down-up-up-\ldots."
This 3D structure is, in fact, rather similar to the
2D one, Fig.~{\ref{fig:order},
found in Sec.~\ref{subsec:ordered}:
observe that spins on the inter-chain bonds in that Figure
follow the same ``4-spin" pattern of two up and two down spins. This
is not a coincidence - in both cases
the classical order can be understood in terms of the well-known
``order-from-disorder" phenomenon,\cite{shender-henley}
induced by either quantum \cite{olegt} or thermal \cite{tsunetsugu}
fluctuations.
This analogy suggests that magnetically ordered state of
GeCu$_2$O$_4$, observed in Ref.~\onlinecite{gecuo},
should be similar to that in the
low-temperature phase of ZnV$_2$O$_4$---clearly more experimental
and theoretical studies of this question are
required. The analogy {\em does not} apply in the $J_\times/J \to 0$ limit
where decoupled $S=1$ chains, although in the
quantum-disordered phase with a finite spin gap
$\Delta \sim 0.4J$,\cite{haldane2} do not break translational symmetry.
This is in contrast with $S=1/2$ chains studied in this paper; the
decoupled limit is characterized by the
crossed-dimer order, Fig.~\ref{fig:pyro3d}, which does break translational
symmetry. Properties of the 3D
phase transitions between quantum-disordered and ordered phases
constitute another interesting theoretical problem
which we leave for future studies.

\section{Conclusions}
\label{sec:conclusions}
The main result of this work consists in the prediction of the novel
VBS phase, the {\sl crossed-dimer} phase, illustrated in
Fig.~\ref{patterns} and Fig.~\ref{fig:pyro3d}.
This new VBS phase arises in the 1D limit of the model as a result of
the frustration-fostered competition
between classical (represented by
the staggered magnetization $\vec{N}$) and quantum (represented by the
staggered dimerization $\epsilon$) ordering tendencies.
Our analysis is based on the careful perturbative implementation of the
well-known $SU(2)$ symmetry of the $g$ matrix field of the WZW
model, which provides
rigorous field-theoretical description of the
low-energy sector of the $S=1/2$ isotropic Heisenberg chain.
This symmetry is made explicit by the OPE
Eqs.~(\ref{ope-r})--(\ref{ope-Je}),
which show transformation properties of
the low-energy ``quantum triad'' $\{\vec{J}_{R/L}, \vec{N}, \epsilon\}$.
As shown in Sec.~\ref{sec:solution}, consistent
implementation of these OPEs requires a careful treatment of the
often-neglected {\sl gradient} terms 
(``non-primary fields'' in the conformal field theory nomenclature, such as $\partial_x \vec{N}$)
which link together quantum fluctuations
of the conserved spin current with those
of the staggered magnetization and dimerization fields.
Once this is done, straightforward
inter-chain perturbation
theory leads to the frustration-generated interaction of
dimerizations
on the crossing chains, Eq.~(\ref{delta-V-epsilon}).

Our finding of the CD-VBS phase in the $J_\times \ll J$ limit of the
checkerboard antiferromagnet eliminates a previously proposed
\cite{sliding} sliding Luttinger liquid phase as the candidate for the
ground state. Like many others, that work \cite{sliding} 
overlooked the crucial role of the gradient terms in the analysis of
the frustrated inter-chain interaction between critical $S=1/2$
Heisenberg chains.

It is also worth pointing out here that our calculation clarifies
previous somewhat inconclusive ``sightings'' of the decoupled-chains phase
\cite{sachdev,toronto} that arise in a widely-used large-$N$ approach 
\cite{read}
to frustrated spin models.  By its very construction, that technique
fails to account, at the leading $N=\infty$ level, for the
fluctuation-generated residual dimer-dimer interaction in the
anisotropic 1D limit (although one does expect finite $1/N$
corrections to the inter-dimer interaction to appear once the
fluctuations of the compact gauge field are accounted
for.\cite{sachdev})

We have also presented a global phase diagram of the CCM
(Sec.~\ref{global}).  Although phenomenological in nature, our
analysis stresses the importance of lattice symmetries in delineating
the {\sl order} of possible direct transitions between various quantum
(CD- and P-VBS) and classical (ordered N\'eel and N\'eel$^*$) phases
of the CCM found in this and previous studies.  We find that most of
such transitions are required to be of the {\sl first-order} type, or
proceed via an intermediate co-existence phases, as illustrated in
Figs.~\ref{fig:global-1} and \ref{fig:global-2}.  This claim concerns
even the relatively well studied P-VBS to N\'eel transition
\cite{sindzingre,brenig,brenig2} and clearly calls for more numerical
(as well as analytical) investigations of this interesting question.

Last but not least, we have also presented a simple but intriguing
extension of the approach to the anisotropic three-dimensional
pyrochlore antiferromagnet, which may be relevant to both $S=1/2$ and
$S=1$ pyrochlore-based magnetic materials.\cite{gecuo,shlee} We hope
that this interesting connection will inspire new experiments in this
exciting area.

\acknowledgments

We would like to thank A. Abanov, P. Azaria, I. Affleck, W. Brenig, F.H.L.
Essler, M.P.A. Fisher, T. Fukui, F.D.M. Haldane, P. Lecheminant, R. Moessner, O.
Motrunich, A.A. Nersesyan, S. Sachdev, P. Sindzingre, O. 
Tchernyshyov, 
H. Tsunetsugu, A.M. Tsvelik, and A. Vishwanath for 
discussions on questions related to
this investigation. We are grateful to O. Tchernyshyov for the help 
with Fig.~\ref{fig:pyro3d}.

We thank Aspen Center for Physics, Kavli Institute for Theoretical
Physics at UC Santa Barbara and Yukawa Institute for Theoretical
Physics at Kyoto University for their hospitality during various
stages of this research.  O.A.S. thanks Research Corporation (award
CC5491) for the partial support of this research.  The work of A.F.
was in part supported by a Grant-in-Aid for Scientific Research
(No.~16GS50219) from Ministry of Education, Culture, Sports, Science
and Technology, Japan.  L.B. was supported by the NSF under grant
DMR-9985255 and by the Packard Foundation.

\appendix
\section{Correlation functions}
\label{cor-functions}

\subsection{Fermionic formulation}

Formally, the WZW model in Eqs.~(\ref{eq:WZW}) and (\ref{eq:sugawara})
describes all the low-energy properties of the $S=1/2$ Heisenberg spin
chains, and is the starting point for a perturbative analysis of
interchain coupling terms.  Unfortunately, as remarked in
Sec.~\ref{subsec:CFT}, this action and
Hamiltonian, while compact and self-contained, are not directly useful
for concrete calculations.  Indeed, in two-dimensional conformal field
theory, analysis often proceeds without {\sl any} action or
Hamiltonian, solely on the basis of algebraic operator product
relations that essentially specify the field content and correlation
functions of the theory.  We will ultimately proceed by using operator
product relations along these lines.  In the crossed chains model,
however, due to the local nature of interactions between perpendicular
chains, we need to pay particular attention to proper short distance
regularization of the theory.  It is therefore desirable to have a more
concrete formulation of the theory.

To do so, we use the well-known phenomena of spin-charge separation in
one-dimensional spin-$1/2$ Dirac fermions.  In particular, using
bosonization, one may show that all degrees of freedom (all states and
operators) of such Dirac
fermions are identical to those of a free scalar ``charge'' boson and
a spin sector described by the WZW $SU(2)_1$ theory of interest to
us. Moreover, the free Dirac fermion Hamiltonian,
\begin{equation}
      \label{eq:Dirac}
      H_d = iv \int\! dx\, \left(\Psi_{L,s}^\dagger \partial_x
      \Psi_{L,s}^{\vphantom\dagger} - \Psi_{R,s}^\dagger \partial_x
      \Psi_{R,s}^{\vphantom\dagger}\right) ,
\end{equation}
can be expressed as the sum of decoupled spin and charge Hamiltonians,
$H_d=H_{WZW}+H_\rho$, with
\begin{equation}
      \label{eq:Hc}
      H_\rho = \frac{v}{2} \int\! dx\, \left[(\partial_x\varphi_\rho)^2 +
      (\partial_x\vartheta_\rho)^2\right],
\end{equation}
where $\varphi_\rho,\vartheta_\rho$ are conjugate ``charge'' boson fields
$[\varphi_\rho(x),\vartheta_\rho(x')]=-i{\rm sign}(x-x')/2$.
The right-moving and left-moving fermion
operators may, if desired, themselves be re-expressed in terms of the
gapless spin and charge fluctuations, which is commonly and
conveniently done with the help of abelian
bosonization.\cite{gnt-book}
We do not, however, require these expressions here.

Because the spin degrees of freedom in
$H_{WZW}$ (described by the $SU(2)$-valued field $g$) are independent
of the charge boson fields, we may safely replace $H_{WZW}$ by $H_d$
at the price of enlarging the physical Hilbert space to include these
charge degrees of freedom, without however affecting the spin physics
in any way.  In particular, regardless of the spin interactions we add
to $H_{WZW}$, the charge sector remains always in its ground state,
the vacuum of Eq.~(\ref{eq:Hc}).

The fermionic formulation provides a convenient way to calculate
while simultaneously regularizing the
short-distance properties of the theory.  This is accomplished by
representing the important operators in the spin sector in terms of
the Dirac fermions.  First, the spin currents may be fully
and simply re-expressed in terms of the Dirac fermions:
\begin{equation}
      \label{eq:JRL-fermions}
      \vec{J}_R= \Psi^\dagger_{R,s} \frac{\vec{\sigma}_{s,s'}}{2}
      \Psi^{\vphantom\dagger}_{R,s'}, \qquad
      \vec{J}_L = \Psi^\dagger_{L,s} \frac{\vec{\sigma}_{s,s'}}{2}
      \Psi^{\vphantom\dagger}_{L,s'}.
\end{equation}
We can thereby construct the uniform magnetization,
$\vec{J}=\vec{J}_R+\vec{J}_L$ in terms of fermions.
Observe that
$\vec{J}$ is invariant under chiral $U(1)$ charge
symmetry
\begin{equation}
\Psi_{R/L,s} \rightarrow e^{i\chi_{R/L}}
\Psi_{R/L,s}
\label{chiral-u1}
\end{equation}
with independent phases
$\chi_R \neq \chi_L$. This invariance implies
independence of
$\vec{J}$ from the charge sector, Eq.~(\ref{eq:Hc}), no matter what its
precise form is.

The second important operator in the spin sector is the staggered
magnetization.  Unlike the uniform magnetization, however, it does not
have a simple expression in terms of the fermions.  Instead, one may
define a fermionic  ``staggered'' (corresponding to the $2k_F$
component for physical electrons in the Tomonaga-Luttinger model) spin
density:
\begin{equation}
\vec{N}_F = \Psi^\dagger_{R,s} \frac{\vec{\sigma}_{s,s'}}{2}
\Psi^{\vphantom\dagger}_{L,s'} + \Psi^\dagger_{L,s}
\frac{\vec{\sigma}_{s,s'}}{2} \Psi^{\vphantom\dagger}_{R,s'}.
\label{N-fermions}
\end{equation}
Using standard bosonization relations, one finds
\begin{equation}
      \label{eq:nnrel}
      \vec{N}_F  =
      \vec{N} \cos \sqrt{2\pi}\varphi_\rho.
\end{equation}
Hence the fermionic staggered spin density, Eq.~(\ref{N-fermions}),
reproduces the desired staggered magnetization, $\vec{N}$, but
multiplied by a factor involving the charge boson $\varphi_\rho$.

Observe that $\vec{N}_F$ is not invariant under the chiral charge
symmetry Eq.~(\ref{chiral-u1})
and this implies that it does couple to
the charge sector of Eq.~(\ref{eq:Dirac}),
as is explicitly stated in
Eq.~(\ref{eq:nnrel}) above. (It is invariant under the diagonal
subgroup
of Eq.~(\ref{chiral-u1}) with $\chi_R=\chi_L$ which only
reflects that it conserves the total charge
of the system, but not
the charges of right- and left- moving chiral sectors independently.)

Similarly, we may define a fermionic staggered dimerization operator,
\begin{equation}
\epsilon_F=\frac{i}{2} \Big(\Psi^\dagger_{R,s}
\Psi^{\vphantom\dagger}_{L,s} - \Psi^\dagger_{L,s}
\Psi^{\vphantom\dagger}_{R,s}\Big)   .
\label{e-fermions}
\end{equation}
Bosonization gives
\begin{equation}
      \label{eq:sdbos}
      \epsilon_F = \epsilon \cos \sqrt{2\pi}\varphi_\rho .
\end{equation}
Like for the staggered magnetization, $\epsilon_F$ defined from the
Dirac fermions reproduces the staggered dimerization operator
$\epsilon$ in the spin sector, but with an undesired multiplicative
charge factor. Here we observe again that this multiplicative
charge
factor follows from the fact that $\epsilon_F$ does not respect
Eq.~(\ref{chiral-u1}),
similarly to $\vec{N}_F$ above.

We will see below that, for our purposes this second factor in
Eqs.~(\ref{eq:nnrel}) and (\ref{eq:sdbos}) is innocuous, so that we may use
$\vec{N}_F,\epsilon_F$ in place of the true $\vec{N},\epsilon$ fields.
We stress this is not generally true, and such a replacement is
possible only under special circumstances we outline below.  For
example, because $\vec{N}_F,\epsilon_F$ contain charge fluctuations,
their correlation functions are {\sl not} the same as those of the
$\vec{N},\epsilon$ operators, and hence the latter cannot be evaluated
using free fermion apparatus.  However, the fermionic operators {\sl
      have the same operator products
with the spin current field $\vec{J}$} as the corresponding spin
operators.
This is because the spin current, as
discussed above, does not depend on charge sector, and as a result the
``charge pieces'' of $\vec{N}_F$ and $\epsilon_F$ stay ``inert'' during
OPE calculations as outlined in Appendix~\ref{a2}.  As it turns out,
such operator products are all we need in order to solve the
problem.  Thus, we can evaluate these crucial operator products using
those of the free fermions.  Furthermore, we can
regularize the problem by adopting a short-distance cutoff for the
Dirac fermions.  This is particularly convenient, as a simple
ultraviolet regularization is available in the Dirac formulation which
keeps SU(2) spin rotation symmetry manifest at all stages.

\subsection{Fermion Green's function.}
\label{app:Green-fermions}

To proceed with our program we need then the Green's function of free
fermions $\Psi_{R/L}$.  As the inter-chain interaction,
Eq.~(\ref{V-ccm-full}), is the sum of local terms over crossings of the
lattice, one needs to be careful about the high-energy regularization
of the continuum theory. We have found that the most natural and
physically appealing regularization is provided by the ``soft-momentum
cutoff" scheme.\cite{haldane,constructive}  In that scheme cutoff
$a$ is introduced via $e^{-a|q|}$ factor in the mode expansion of
collective bosonic fields.  In going from the momentum to the
coordinate space, cutoff dependence on $a$ transforms into that on
$\alpha\equiv a/v$ in the temporal ($\tau$) direction.  Adopting this
scheme, one finds that the zero temperature ($T=0$) Green's function
for the right-moving fermions on the same chain (chain index is
omitted for brevity) is given by
\begin{eqnarray}
F_R(x,\tau) &=&
-\langle \hat{T} \Psi_{R,s}(x,\tau) \Psi_{R,s}^+(0,0)\rangle
\nonumber\\
&=&
-\frac{1}{2\pi v(\tau - ix/v + \alpha \sigma_\tau)},
\label{f-r}
\end{eqnarray}
where $\alpha= a/v$ is the ultraviolet cutoff in temporal direction and
$\sigma_\tau = \text{sign}(\tau)=\Theta(\tau) - \Theta(-\tau)$ is the
sign-function. For the left-moving ones
\begin{eqnarray}
F_L(x,\tau) &=& F_R^*(x,\tau) = F_R(-x,\tau) = - F_R(x,-\tau)
\nonumber\\
&=& -\frac{1}{2\pi v(\tau + ix/v + \alpha \sigma_\tau)}.
\label{f-l}
\end{eqnarray}
It is often convenient to introduce complex coordinates
\begin{equation}
z=\tau + ix/v , \quad
\bar{z} = \tau - ix/v
\label{z-coordinate}
\end{equation}
in terms of which
\begin{equation}
F_R(\bar{z}) = -\frac{1}{2\pi v \bar{z}} , \quad
F_L(z)=-\frac{1}{2\pi v z}.
\label{f-z}
\end{equation}
It is important to keep in mind, however, that expressions of
Eqs.~(\ref{f-r}) and (\ref{f-l}) are ``superior'' to Eq.~(\ref{f-z})
in that they contain
an explicit cutoff prescription.

\section{Derivation of OPE Eq.~(\ref{ope-JN})}
\label{a2}

We need Eqs.~(\ref{eq:JRL-fermions}) and (\ref{N-fermions})
and the single-particle
Green's function, Eqs.~(\ref{f-r}) and (\ref{f-l}).
Consider the product of $J_R^a(x_1,\tau_1)$ and $N_F^b(x_2,\tau_2)$
at nearby points
(such that $|z_1-z_2| \sim \alpha$ with $z_i=\tau_i+ix_i/v$).
To shorten notations, we denote $(x_1,\tau_1)={\bf 1}$, etc.
Now use Wick's theorem for {\sl free} fermions, as appropriate for
Eq.~(\ref{eq:Dirac}),
as well as the identity $\sigma^a \sigma^b =
\delta^{ab} + i\epsilon^{abc} \sigma^c$,
to reduce the product of
four $\Psi$'s to that of two of them and $F_R$
\begin{eqnarray}
&&J_R^a(\mathbf{1}) N_F^b(\mathbf{2})
= -\frac{1}{4} F_R(\mathbf{1}-\mathbf{2})
\nonumber\\
&&\times\Big( i \epsilon^{abc} \sigma^c_{s,s'}
[\Psi^+_{R,s}(\mathbf{1}) \Psi_{L,s'}(\mathbf{2})
    + \Psi^+_{L,s}(\mathbf{1}) \Psi_{R,s'}(\mathbf{2})]
\nonumber\\
&&\quad{}
+\delta^{ab} [\Psi^+_{R,s}(\mathbf{1}) \Psi_{L,s}(\mathbf{2})
- \Psi^+_{L,s}(\mathbf{1}) \Psi_{R,s}(\mathbf{2})] \Big),
\label{fuse}
\end{eqnarray}
where summation over repeated spin indices $s$ and $s'$ is assumed.
Fusing $(x_1,\tau_1)$ and $(x_2,\tau_2)$ points, one finds that
the coefficient
of $\epsilon^{abc}$ is just $2 N_F^c(\mathbf{2})$,
whereas that of $\delta^{ab}$ is $-2i \epsilon_F(\mathbf{2})$,
the staggered dimerization operator; see Eq.~(\ref{e-fermions}).
Hence
\begin{equation}
J_R^a(\mathbf{1}) N_F^b(\mathbf{2})=
\frac{ i \epsilon^{abc} N_F^c(\mathbf{2})
    - i \delta^{ab}\epsilon_F(\mathbf{2})}
{4\pi v(\bar{z}_1-\bar{z}_2+\alpha\sigma_{\tau_1-\tau_2})}.
\label{ope-JN-app}
\end{equation}
The second OPE, between $\vec{J}_L$ and $\vec{N}_F$,
is obtained by replacing $F_R \rightarrow F_L$ and {\em changing the
sign} of the last term in
Eq.~(\ref{ope-JN-app}) as is readily verified by the explicit calculation.
Thus
\begin{equation}
J_L^a(\mathbf{1}) N_F^b(\mathbf{2})=
\frac{i \epsilon^{abc} N_F^c(\mathbf{2})
    + i \delta^{ab}\epsilon_F(\mathbf{2})}
{4\pi v(z_1-z_2+\alpha\sigma_{\tau_1-\tau_2})}.
\label{ope-l-app}
\end{equation}

Observe now that the multiplicative charge factor
$\cos\sqrt{2\pi} \varphi_\rho$ from Eqs.~(\ref{eq:nnrel}) and
(\ref{eq:sdbos})
appears on both sides of Eqs.~(\ref{ope-JN-app}) and (\ref{ope-l-app})
above.
Dividing by it,
we obtain the OPE quoted
in Eq.~(\ref{ope-JN})
\begin{eqnarray}
&&J^a_R(\bar{z}) N^b(w,\bar{w}) = \frac{ i
\epsilon^{abc} N^c(w,\bar{w}) - i
\delta^{ab}\epsilon(w,\bar{w})}
{4\pi v
(\bar{z}-\bar{w})},\nonumber\\
&&J^a_L(z) N^b(w,\bar{w}) = \frac{ i
\epsilon^{abc} N^c(w,\bar{w}) + i
\delta^{ab}\epsilon(w,\bar{w})}
{4\pi v
(z-w)}.
\qquad
\label{app:JN-ope}
\end{eqnarray}

It is instructive to repeat the same calculation  using abelian
bosonization.
Consider, for example, $J^z_R(x_1,\tau_1)
N^x(x_2,\tau_2)$.
Abelian bosonization tells us that
\begin{eqnarray}
J^z_R({\bf 1}) &=& \frac{1}{\sqrt{2\pi}} \partial_{x_1}
\Phi_{R,\sigma}({\bf 1}),
\label{J^z-ab}
\\
N^x({\bf 2}) &=& \frac{\lambda}{4\pi a}
\left(
e^{-i\sqrt{2\pi}[\Phi_{R,\sigma}({\bf 1})
    - \Phi_{L,\sigma}({\bf 1})]} + \mathrm{H.c.}\right),
\qquad
\label{N^x-ab}
\end{eqnarray}
where $\Phi_{R/L,\sigma}$ are the {\em spin} components of chiral
right/left bosons,
and the scale factor $\lambda= \langle
\cos\sqrt{2\pi} \varphi_\rho \rangle_\rho$
is the expectation value
of the charge field. The brackets $\langle ...\rangle_\rho$ denote
average with respect to
the charge Hamiltonian $H'_\rho$ which
includes now four-fermion {\sl umklapp} term responsible for
the
opening of the charge gap (as a result of which $\lambda \neq 0$),
see Appendix A of Ref.~\onlinecite{shelton-ladder} for more details.

Thus charge fluctuations
are explicitly separated from spin ones in the manipulations to follow.
To fuse Eq.~(\ref{J^z-ab}) and Eq.~(\ref{N^x-ab}), we need
\begin{equation}
\Phi_{R,\sigma}({\bf 1}) e^{i \beta
\Phi_{R,\sigma}({\bf 2})} =
i\beta
\langle \hat{T} \Phi_{R,\sigma}({\bf 1}) \Phi_{R,\sigma}({\bf 2})\rangle
e^{i \beta \Phi_{R,\sigma}({\bf 2})},
\label{Phi}
\end{equation}
which is obtained by expanding exponential on the left-hand side of
Eq.~(\ref{Phi}),
pairing $\Phi_{R,\sigma}({\bf 1})$ with $\Phi_{R,\sigma}({\bf 2})$
in all possible ways (Wick's theorem for free chiral spin bosons),
and collecting the rest  of the series back into
exponential.
Using Eq.~(\ref{Phi}) and the fact that right and left bosons are
independent from each other, we find
\begin{equation}
J^z_R({\bf 1}) N^x({\bf 2}) =
\partial_{x_1}
\langle \hat{T}
\Phi_{R,\sigma}({\bf 1}) \Phi_{R,\sigma}({\bf 2})
\rangle N^y({\bf 2}).
\end{equation}
Finally,
\begin{equation}
\langle \hat{T} \Phi_{R,\sigma}({\bf 1}) \Phi_{R,\sigma}({\bf 2})\rangle
= - \frac{1}{4\pi}
\ln[\alpha + \sigma_{\tau_1-\tau_2}(\bar{z}_1-\bar{z}_2)]
\end{equation}
for free chiral bosons.
Hence
\begin{equation}
J^z_R({\bf 1}) N^x({\bf 2}) =
\frac{i N^y({\bf 2})}
{4\pi v(\bar{z}_1 - \bar{z}_2 + \alpha \sigma_{\tau_1 - \tau_2})},
\end{equation}
which is just one component of Eq.~(\ref{ope-JN-app}).
To get $\delta^{ab}$ term of Eq.~(\ref{ope-JN-app}) one has to consider
explicitly OPE of, say, $J^z$ and $N^z$ fields.


\begin{thebibliography}{99}
\bibitem{paradigm} R.R.P. Singh, O.A. Starykh, and P.J. Freitas,
J.\ Appl.\ Phys.\ {\bf 83}, 7387 (1998).
\bibitem{canals} B. Canals, \prb {\bf 65}, 184408 (2002).
\bibitem{olegt} O. Tchernyshyov, O.A. Starykh, R. Moessner,
    and A.G. Abanov,
    \prb {\bf 68}, 144422 (2003).
\bibitem{sachdev} C.H. Chung, J.B. Marston, and S. Sachdev,
    \prb {\bf 64}, 134407 (2001).
\bibitem{moessner} R. Moessner, O. Tchernyshyov, and S.L. Sondhi,
    J.\ Stat.\ Phys.\ \textbf{116}, 755 (2004); cond-mat/0106286.
\bibitem{toronto} J.S. Bernier, C.H. Chung, Y.B. Kim, and S. Sachdev,
    \prb \textbf{69}, 214427 (2004).
\bibitem{hermele} M. Hermele, M.P.A. Fisher, and L. Balents,
    \prb {\bf 69}, 064404 (2004).
\bibitem{sliding} O.A. Starykh, R.R.P. Singh, and G.C. Levine,
    \prl {\bf 88}, 167203 (2002).
\bibitem{chalker} S.E. Palmer and J.T. Chalker,
    \prb {\bf 64}, 094412 (2001).
\bibitem{fouet} J.B. Fouet, M. Mambrini, P. Sindzingre, and
    C. Lhuillier, \prb {\bf 67}, 054411 (2003).
\bibitem{sindzingre} P. Sindzingre, J.B. Fouet, and C. Lhuillier,
    \prb {\bf 66}, 174421 (2002).
\bibitem{brenig} W. Brenig and A. Honecker, \prb {\bf 65}, 140407 (2002).
\bibitem{altman} E. Berg, E. Altman, and A. Auerbach, \prl {\bf 90},
    147204 (2003).
\bibitem{brenig2} W. Brenig and M. Grzeschik, \prb {\bf 69}, 064420 (2004).
\bibitem{oleg-leon} O.A. Starykh and L. Balents,
    \prl \textbf{93}, 127202 (2004).
\bibitem{rg-affleck} I. Affleck and B.I. Halperin,
    J.\ Phys.\ A: Math.\ Gen.\ {\bf 29}, 2627 (1996).
\bibitem{schulz} H. Schulz, \prl {\bf 77}, 2790 (1996).
\bibitem{allen-essler-ners} D. Allen, F.H.L. Essler, and A.A. Nersesyan,
    \prb {\bf 61}, 8871 (2000).
\bibitem{kim} E.H. Kim, G. F\'ath, J. S\'olyom, and D.J. Scalapino,
    \prb {\bf 62}, 14965 (2000).
\bibitem{NT} A.A. Nersesyan and A.M. Tsvelik,
    \prb {\bf 67}, 024422 (2003).
\bibitem{wzw} I. Affleck and F.D.M. Haldane, \prb {\bf 36}, 5291 
(1987).
\bibitem{itzykson} C. Itzykson and J.-M. Drouffe,
{\em Statistical Field Theory}, vol. 2,
Appendix 9C (Cambridge
University Press, Cambridge, UK, 1989).
\bibitem{gnt-book} A.O. Gogolin, A.A. Nersesyan, and A.M. Tsvelik,
    {\em Bosonization and Strongly Correlated Systems}
    (Cambridge University Press, Cambridge, UK, 
1998).
\bibitem{eggert} S. Eggert and I. Affleck, \prb {\bf 46}, 
10866 (1992).
\bibitem{lin} H.-H. Lin, L. Balents, and M.P.A. Fisher,
    \prb {\bf 56}, 6569 (1997).
\bibitem{affleck-thanks} I. Affleck, private 
communication.
\bibitem{shelton-ladder} D.G. Shelton, A.A. Nersesyan 
and A.M. Tsvelik,
    \prb {\bf 53}, 8521 (1996).
\bibitem{luk} S. Lukyanov and A. Zamolodchikov,
    Nucl.\ Phys.\ B {\bf 493}, 571 (1997).
\bibitem{starykh96} O.A. Starykh, M.E. Zhitomirsky, D.I. Khomskii,
   R.R.P. Singh, and K. Ueda, \prl \textbf{77}, 2558 (1996).
\bibitem{zhitomirsky} M.E. Zhitomirsky and K. Ueda,
   \prb \textbf{54}, 9047 (1996).
\bibitem{lauchli} A. L\"auchli, S. Wessel, and M. Sigrist,
    \prb \textbf{66}, 014401 (2002).
\bibitem{brenig-discussion} We thank
W. Brenig for the discussions on this 
point.
\bibitem{landau-statmech} L.D. Landau and E.M. Lifshits,
   \textit{Statistical Physics, Part 1, 3rd Edition}
   (Butterworth-Heinemann, New York, 2003).
\bibitem{dcp} T. Senthil, L. Balents, S. Sachdev, A. Vishwanath,
   and M.P.A. Fisher, \prb {\bf 70}, 144407 
(2004).
\bibitem{haldane88} F.D.M. Haldane, \prl {\bf 61}, 1029 
(1988).
\bibitem{motrunichprivate} O.I. Motrunich, private communication.
\bibitem{sachdev-park} S. Sachdev and K. Park,
   Ann.\ Phys.\ (N.Y.) {\bf 298}, 58 (2002).
\bibitem{gecuo} T. Yamada, Z. Hiroi, M. Takano, M. Nohara, and H. Takagi,
    J.\ Phys.\ Soc.\ Jpn.\ {\bf 69}, 1477 (2000).
\bibitem{znvo} Y. 
Ueda, N. Fujiwara, and H. Yasuoka,
    J.\ Phys.\ Soc.\ Jpn.\ {\bf 66}, 778 (1997).
\bibitem{shlee} S.-H. Lee, D. Louca, H. Ueda, S. Park, T.J. Sato,
    M. Isobe, Y. Ueda, S. Rosenkranz, P. Zschack, J. Iniguez,
    Y. Qiu, and R. Osborn, \prl {\bf 93}, 156407 (2004).
\bibitem{tsunetsugu} Y. Motome and H. Tsunetsugu,
    \prb \textbf{70}, 184427 (2004);
    H. Tsunetsugu and Y. Motome,
    \prb {\bf 68}, 060405 (2003).
\bibitem{shender-henley} E.F. Shender,
Sov.\ Phys.\ JETP {\bf 56}, 178 (1982);
    C.L. Henley, \prl {\bf 62}, 2056 (1989).
\bibitem{haldane2} F.D.M. Haldane,
   Phys.\ Lett.\ \textbf{93A}, 464 (1983);
    \prl \textbf{50}, 1153 (1983).
\bibitem{read} N. Read and S. Sachdev, \prb {\bf 42}, 4568 (1990).
\bibitem{haldane} F.D.M. Haldane, J.\ Phys.\ C \textbf{14}, 2585 (1980).
\bibitem{constructive} J. von Delft and H. Schoeller,
    Ann.\ Phys.\ (Leipzig) {\bf 7}, 225 (1998).

\end{thebibliography}
\end{document}